# Improving Disease Classification Performance and Explainability of Deep Learning Models in Radiology with Heatmap Generators

Akino Watanabe[1], Sara Ketabi[2,3], Khashayar (Ernest) Namdar[2,5,7], and Farzad Khalvati[2,3,4,5,6,7]

[1]Engineering Science, University of Toronto, Toronto, ON, Canada
[2]The Hospital for Sick Children (SickKids), Toronto, ON, Canada
[3]Department of Mechanical and Industrial Engineering, University of Toronto, Toronto, ON, Canada
[4]Department of Medical Imaging, University of Toronto, Toronto, ON, Canada
[5]Institute of Medical Science, University of Toronto, Toronto, ON, Canada
[6]Department of Computer Science, University of Toronto, Toronto, ON, Canada
[7]Vector Institute, Toronto, ON, Canada

## Abstract

As deep learning is widely used in the radiology field, the explainability of such models is increasingly becoming essential to gain clinicians' trust when using the models for diagnosis. In this research, three experiment sets were conducted with a U-Net architecture to improve the classification performance while enhancing the heatmaps corresponding to the model's focus through incorporating heatmap generators during training. All of the experiments used the dataset that contained chest radiographs, associated labels from one of the three conditions ("normal", "congestive heart failure (CHF)", and "pneumonia"), and numerical information regarding a radiologist's eye-gaze coordinates on the images. The paper (A. Karargyris and Moradi, 2021) that introduced this dataset developed a U-Net model, which was treated as the baseline model for this research, to show how the eye-gaze data can be used in multi-modal training for explainability improvement. To compare the classification performances, the 95% confidence intervals (CI) of the area under the receiver operating characteristic curve (AUC) were measured. The best method achieved an AUC of 0.913 (CI: 0.860-0.966). The greatest improvements were for the "pneumonia" and "CHF" classes, which the baseline model struggled most to classify, resulting in AUCs of 0.859 (CI: 0.732-0.957) and 0.962 (CI: 0.933-0.989), respectively. The proposed method's decoder was also able to produce probability masks that highlight the determining image parts in model classifications, similarly as the radiologist's eye-gaze data. Hence, this work showed that incorporating heatmap generators and eye-gaze information into training can simultaneously improve disease classification and provide explainable visuals that align well with how the radiologist viewed the chest radiographs when making diagnosis.

## 1. Introduction

Complex deep learning models have been more recently incorporated into clinical radiology practices and have given promising results to assist radiologists in identifying and classifying various diseases and abnormalities (P. Linardatos and Kotsiantis, 2020), such as pneumonia and congestive heart failure (CHF). However, it is quite difficult for humans, including radiologists, to understand how such deep learning models arrived at their predictions (A. Singh and Lakshminarayanan, 2020). Unlike linear regression or support vector machines (SVMs) that have fewer-dimensional classification boundaries, which are easier to understand and visualize, deep learning algorithms are often referred to as black-box algorithms (A. Singh and Lakshminarayanan, 2020) because of their computational complexity and the fact that we are often unable to easily observe the decision boundaries generated by those models (P. Linardatos and Kotsiantis, 2020; P. P. Angelov and Atkinson, 2021; J. D. Fuhrman and Giger, 2021). In fact, performance and explainability have often been traded-off, and models with better performance tend to be worse in terms of explainability and vice versa (P. P. Angelov and Atkinson, 2021; A. Holzinger and Kell, 2017; J. Amann and Madai, 2020).

Hence, the explainability of AI models that shows which part of the model's inputs or what kind of information the models focused on when making predictions (A. Singh and Lakshminarayanan, 2020) is crucial for the radiology field to gain and increase radiologists', patients', and regulators' trust in the use of AI models for diagnosis (A. Holzinger and Kell, 2017). The explainability aspect of the models also helps to verify the model's conclusions and to identify models' biases (Reyes et al., 2020; J. Amann and Madai, 2020; Fernandez-Quilez, 2022). A model with improved classification and enhanced explainability components will be able to assist the radiologists by making the diagnosis process more efficient and minimizing the risk of making mistakes (Reyes et al., 2020). With the increase in explainability, the model can also be widely used at medical facilities that perform Chest X-Ray (CXR) imaging where there may not be enough radiological expertise to identify and classify diseases.

The methods to establish the explainability of the models can involve different data types (P. Linardatos and Kotsiantis, 2020), such as mathematical computations, texts, or visualization with heatmaps or saliency maps. In the radiology field, CXR images are one of the essential components for diagnosing abnormalities or conditions that affect chest and nearby organs, and many AI models for radiology often use the chest radiographs as one of their inputs (E. Sogancioglu and Murphya, 2021; Reyes et al., 2020). Thus, the visualization aspect of explainability on images with heatmaps is an impactful tool to convey which part of the radiograph the model observed closely for illness classification (P. Linardatos and Kotsiantis, 2020; R. Shad and Hiesinger, 2021).

Currently, there are several studies on using deep learning models for disease classification that also involve heatmap generation to visualize where the model's focus was on the given input CXR image (E. Sogancioglu and Murphya, 2021; W. Kusakunniran, 2021). Nevertheless, there are many limitations in such research. For example, many publicly available datasets that such studies use contain erroneous class labels because the labels were often extracted from text reports using natural language processing (NLP) models (A. Karargyris and Moradi, 2021; E. Sogancioglu and Murphya, 2021), and the state-of-the-art NLP models still cannot achieve the 100% accuracy on texts interpretation. Additionally, many of the models use only CXR images and disease class labels for training, but many diseases can only be classified through using other information the CXR images cannot necessarily provide, such as symptoms, clinical signs, patients' history, and results from blood tests (E. Sogancioglu and Murphya, 2021). Lastly, many of the studied models do not consider the methods the radiologists usually take to analyze the CXR images, such as the way they view the chest radiographs to make diagnosis (A. Karargyris and Moradi, 2021). It has been shown that integrating eye-gaze information improves AI models' classification performance and has been validated that eye-gaze data contains valuable information related to focusing on important input features (Y. Rong and Kasneci, 2021).

To increase the explainability of AI models while mitigating such dataset and labeling limitations, this work proposes and tests three different methods to observe whether incorporating radiologists' eye-gaze information on CXR images and various heatmap generators in addition to CXR images and class labels into a convolutional-based U-Net (O. Ronneberger and Brox, 2015) model and its training can improve disease classification while enhancing explainability. This research used a dataset (A. Karargyris and Moradi, 2021) that contains not only the CXR images and corresponding class labels (which are "normal", "CHF", and "pneumonia"), but it also contains a radiologist's eye-gaze coordinates information received when the radiologist was viewing the CXR image to perform diagnosis.

Although the proposed models, which uniquely used the heatmap generators' outputs and the radiologist's eye-gaze data during models' training, mostly had similar overall average classification performances as the baseline model that did not use heatmap generators in the training process, there were greater improvements in classifying the "CHF" and "pneumonia" classes, which were the classes the baseline model struggled to correctly classify. Moreover, one of the proposed models had a superior improvement in both the overall average classification performance and the performance on "CHF" and "pneumonia" classes.

2. Background

This section reviews the notion of the "explainability of AI" used in this study. Furthermore, two prominent paths of incorporating explainability into AI for radiology are specified, which are visualizing with saliency maps and training deep learning models to directly generate attention maps as one of their outputs.

2.1 Explainability of AI in Radiology

The concept of "explainability of AI" is often defined as the ability of someone to understand and see which extracted features of the input data directly used by an AI model actually contributed to the model's predictions (A. Singh and Lakshminarayanan, 2020; P. Linardatos and Kotsiantis, 2020; A. Holzinger and Kell, 2017). It also helps to identify models' biases and understand how the model achieved its predictions (Reyes et al., 2020). Methods for explainability of AI should ensure that models did not operate on unrelated features (K. Preechakul and Chuangsuwanich, 2022).

There are two major key methods to showcase the explainability aspects of AI models for the radiology field when using chest radiographs as a part of a model's training. Firstly, various saliency maps (also known as attention-maps, heatmaps, or sensitivity maps) generation methods can be incorporated within, or after training a deep learning model that highlight the areas of the input chest radiographs the model's parameters focused on when making predictions. Secondly, there are deep learning models that were developed to simultaneously operate disease classification and heatmap generation using image segmentation.

2.1.1 Various Saliency Maps and Attention Map Generators

Saliency maps show the parts of the input image that contributed most to the model's output predictions (Salehi, 2020; Z. Salahuddina and Lambin, 2022). Some of the post-hoc, gradient-based methods that can be used to obtain the saliency maps are class activation maps (CAM), Gradient-weighted Class Activation Mapping (Grad-CAM) (R. Selvaraju and Batra, 2016), deconvNet (Zeiler and Fergus, 2013), back-propagation, and guided back-propagation (J. T. Springenberg and Riedmiller, 2015), (Reyes et al., 2020). As can be seen in Figure 1, each of the generators produces visually different heatmaps that highlight the section of the image that corresponds to the class that a model predicted. All of them are variations of deconvolution and back-propagation methods (E. Mohamed and Howells, 2022), where deconvolution attempts to recreate the input image from the activations of the model's layer, while back-propagation is how a model's weights change during training time to decrease its loss and is used to find the relevance of the input pixels to the output predictions based on how the gradients were assigned to those pixels (Salehi, 2020; Draelos, 2019). The generated heatmaps are the visualization of such computations or gradients. Gradients with a larger magnitude signify that the corresponding pixels have more influence in the specific classification of the image.

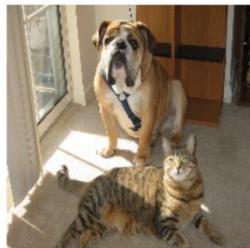

*Figure 1: The outputs of various heatmap generators for the given image (Nakashima, 2020)*

Procedures for deconvNets and guided back-propagation are quite similar where the difference between them is the way each backpropagates through ReLU non-linearity as shown in Figure 2. DeconvNets run the usual back-propagation by using transpose convolution and undoing pooling operations, but they back-propagate only the positive error signals for the ReLU activation (Draelos, 2019).

Guided back-propagation maps are produced through the combination of the deconvolution and back-propagation methods. The deconvolution part shows which pixels contributed positively to the model's output through selectively back-propagating the positive component of the gradients between the input and the output of the models; meanwhile the back-propagation part restricts the model to consider only the positive inputs, which can result in more zeros in the outputs than DeconvNets (Draelos, 2019), and hence, guided back-propagation maps tend to have higher resolution.

CAM (B. Zhou and Torralba, 2015) is another visualization map, which is produced using a global average pooling layer in CNN models, where the pooling layer reduces each feature map into one number, so that the weights connecting the pooling layer and the final classification layer encodes the contribution and the importance of each feature map to the final class prediction. Figure 3 illustrates how CAM works and what the output heatmaps may resemble.

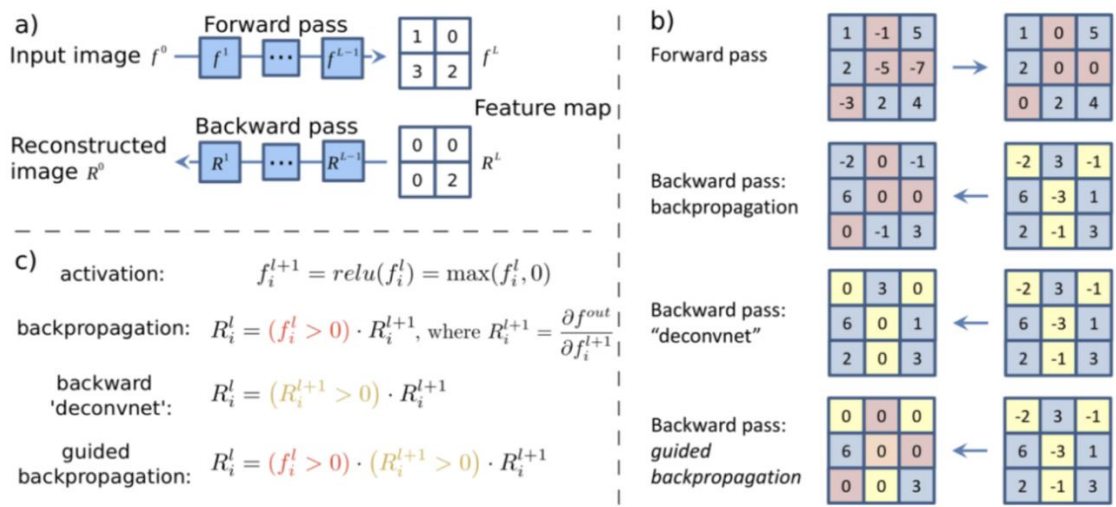

*Figure 2: Computation differences between deconvNets and guided back-propagation (J. T. Springenberg and Riedmiller, 2015)*

There are several modifications and improvements made on CAM for image analysis and computer vision tasks. For example, Pyramid Localization Network (PYLON) (K. Preechakul and Chuangsuwanich, 2022) was developed to produce higher resolution of CAM heatmaps with greater preciseness using Pyramid Attention mechanism and upsampling blocks.

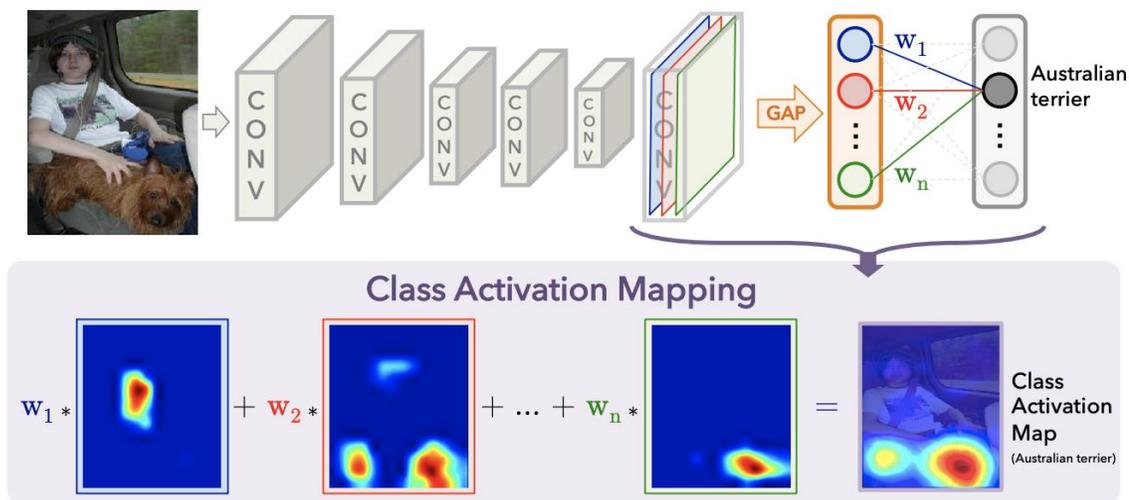

*Figure 3: Decomposition of CAM Computation (B. Zhou and Torralba, 2015)*

Because CAM can only be produced using a specific set of CNN models, such as those without fully connected layers, Grad-CAM was proposed as a generalization of CAM to eliminate the necessity of trade-off between model accuracy and explainability and not to require model retraining (P. Linardatos and Kotsiantis, 2020). It can be applied to many variations of CNN models, including those with fully connected layers (R. Selvaraju and Batra, 2016; P. P. Angelov and Atkinson, 2021; P. Linardatos and Kotsiantis, 2020). Grad-CAM focuses on the gradients flowing into the last convolutional layer of the model and assigns importance scores to each neuron to generate localization maps. The importance score, which is the contribution of a specific feature map to the model's output, is computed by finding the gradient values for a specific class with respect to the activation map of a convolutional layer and using global average pooling on the gradients. This results in a coarse heatmap that is the same size as the convolutional feature map and will only consider features that have a positive influence on the specific class.

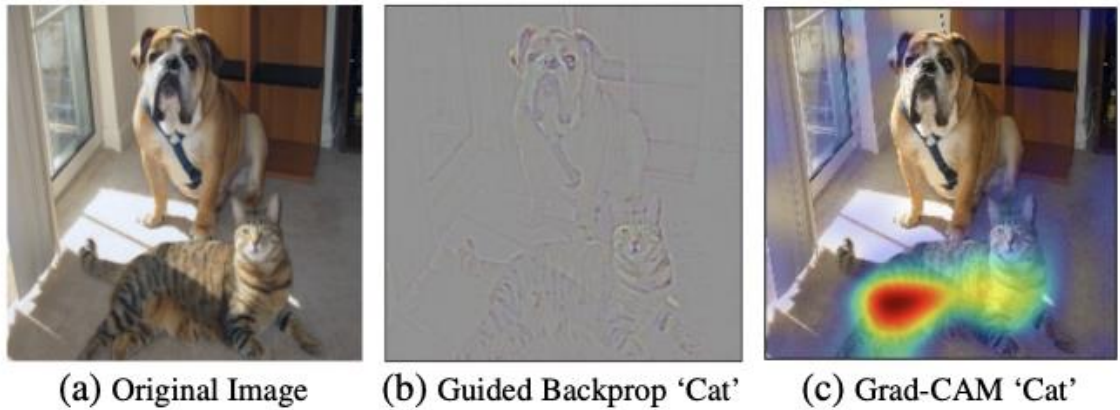

*Figure 4: Comparison of guided back-propagation and Grad-CAM (R. Selvaraju and Batra, 2016)*

As a comparison, Figure 4 is an example of what the grad-CAM and the guided back-propagation maps look like, given an input image shown on the left.

There are multiple variations of grad-CAM, including guided grad-CAM and GradCAM++. Guided grad-CAM is an element-wise multiplication of guided back-propagation and grad-CAM, which results in high-resolution, class-discriminative maps. GradCAM++ (A. Chattopadhyay and Balasubramanian, 2017) focuses on the weighted average of the positive partial derivatives of the last convolutional layer's feature maps with respect to a specific class for a better object localization and recognizing multiple class objects in a single image.

SmoothGrad is a method that reduces noise in the output saliency maps by first adding various noise into an input image, and second producing look-alike images, and third using those images to generate many saliency maps that can be averaged to produce one saliency map with less noise. It is addressed in response to the possibility that the noise in the saliency maps is due to irrelevant, local fluctuations in partial derivatives. Some outputs of using smoothGrad with different sample sizes when averaging can be seen in Figure 5, and outputs indicating the effect of the noise level using smoothGrad is depicted in Figure 6. It has been shown that combining smoothGrad and guided back-propagation methods can produce more visually coherent maps (D. Smilkov and Wattenberg, 2017), as can be seen in the leftmost images in Figure 7.

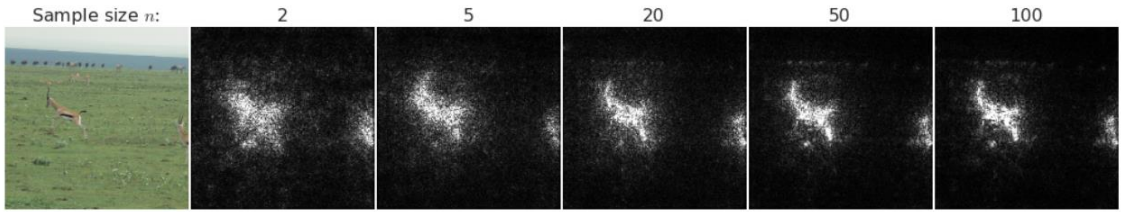

*Figure 5: Effects of smoothGrad with different sample sizes given an input image (D. Smilkov and Wattenberg, 2017)*

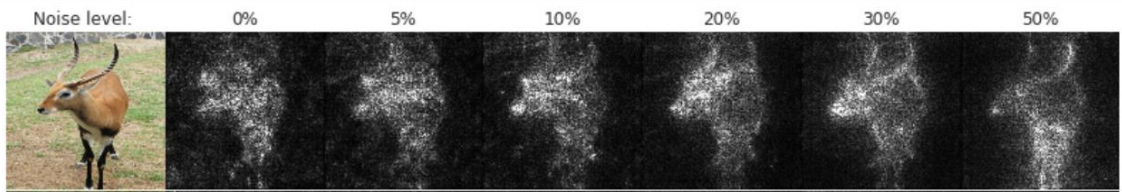

*Figure 6: Effects of smoothGrad with different noise levels given an input image (D. Smilkov and Wattenberg, 2017)*

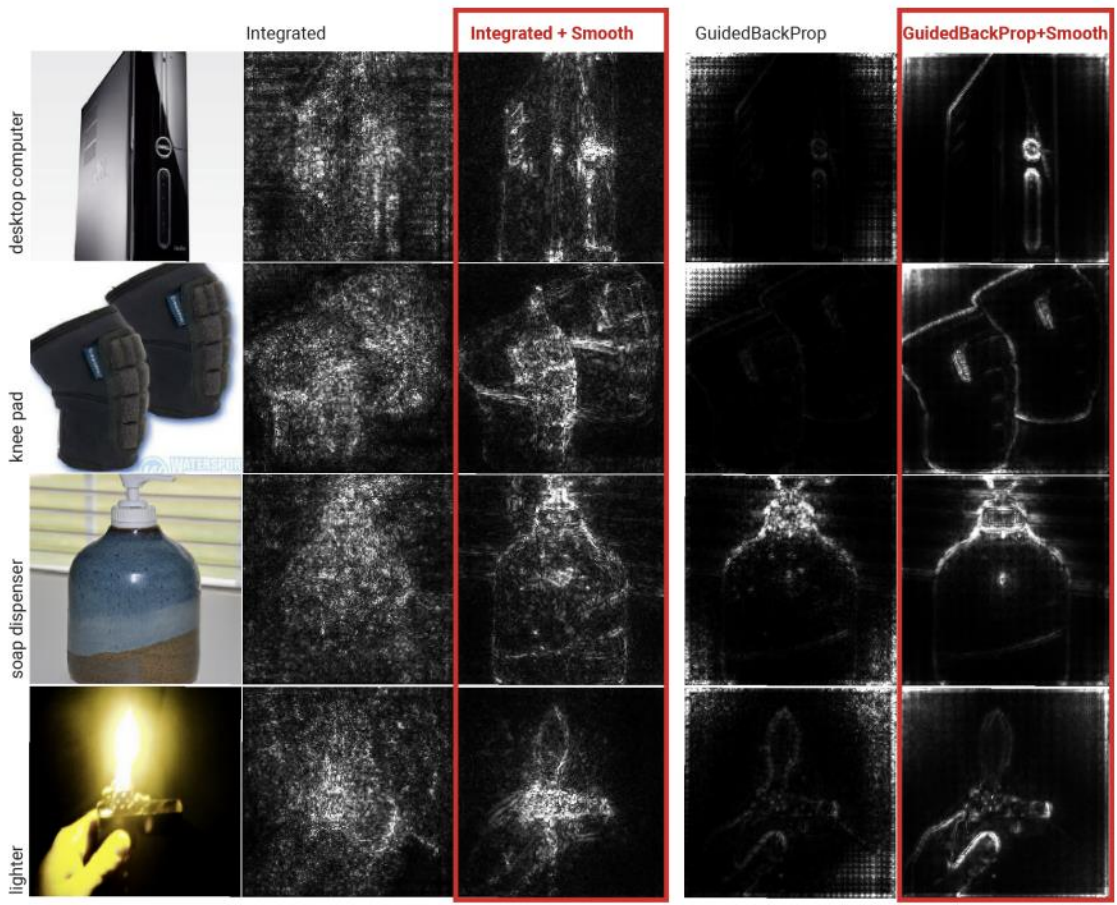

*Figure 7: Effects of combining smoothGrad and guided back-propagation on given input images (D. Smilkov and Wattenberg, 2017)*

DeepLIFT (Deep Learning Important FeaTures) (A. Shrikumar, 2019) also uses back-propagation method for explainability. Specifically, DeepLIFT employs "difference from reference" concept for neurons' activations to determine importance scores for each input through back-propagating the model once.

Unlike the use of gradient-methods and back-propagation related methods for explainability, LIME (Local Interpretable Model-Agnostic Explanations) (M. R. Ribeiro and Guestrin, 2016) creates an interpretable model through approximating the explanations locally around classifiers and providing a method that chooses several combinations of example inputs and corresponding explanations to address why the model should be trusted. LIME can give explanations to model predictions for any classifier or regressor.

Although there is much debate about the true effectiveness of saliency map generation, the maps give insights into how a model arrives at its prediction output, and they are often used post-training to visualize the model's attention.

2.2 Deep learning models as heatmap generators and classifiers

As opposed to the above heatmap generating methods that can be inserted during or after training deep learning models, there are several studies involving constructing deep learning models or pipelines that focus on generating heatmaps as part of the model's outputs and on operating localization tasks. One of the more widely used model architecture for image segmentation tasks is a U-Net (O. Ronneberger and Brox, 2015), as it consists of several convolutional layers in both downsampling and upsampling paths while using skip connections between the layers in those paths to maintain high resolution and fine details of the model's inputs throughout its training time (E. Sogancioglu and Murphya, 2021). U-Net's decoder can produce attention maps or recreate the input images depending on how the model is trained, and hence U-Net architecture is often desirable for joint image classification and localization tasks. Other architectures that can be used for image localization include YOLO, mask R-CNN, and faster R-CNN (E. Sogancioglu and Murphya, 2021).

To tackle the localization problem in the case where the annotated datasets are not large, the use of the combinations of different techniques, such as a classifier and a localizer that produces bounding boxes and probability heatmaps (E. Pescea and Montana, 2019), is shown to be effective. Some notable methods include Weakly Supervised Learning proposed in (O. Viniavskyi and Dobosevych, 2020; A. Chaudhry and Torr, 2017; Y. Wei and Yan, 2018), Self-Transfer Learning (Hwang and Kim, 2016), and other models, such as those introduced in (P. Rajpurkar, 2017).

For example, a proposed weakly supervised leaning method in (O. Viniavskyi and Dobosevych, 2020) consists of a three-stage network for disease localization that first generates class activation maps, and second, feeds the maps into a network that outputs pseudo labels, and finally uses the pseudo labels for image segmentation.

As another example, a self-transfer learning framework introduced in (Hwang and Kim, 2016) consists of three components (convolutional layers, fully connected classification layers, and localization layers) where the goal is to perform image localization. The classification branch and localization branch of the model are trained simultaneously with two losses. To ensure that the localization component does not stray away, the weights on the classification loss decrease through the training process while the weights on the localization loss increase.

Although saliency maps are often produced and analyzed post-training, there are several architectures that incorporate attention maps and map generating mechanisms in model training to improve explainaibility and classification or segmentation performance, such as those introduced in (M. Kazemimoghadam and Gu, 2021; A. A. Ismail and Bravo, 2021; K. Li and Fu, 2018; L. Wang and Metaxas, 2019; A. Ross and Doshi-Velez, 2017).

(M. Kazemimoghadam and Gu, 2021) incorporated saliency maps as one of the inputs to multiple U-Net models for post-operative tumor bed volume segmentation in CT images. Specifically, the saliency maps are encoded using markers in CT images, which will be combined with CT images to guide the model to focus on higher intensity values in the saliency maps to extract relevant features for a more accurate segmentation. Ultimately, voting occurs among multiple U-Net models to produce a final segmentation prediction.

Unlike how (M. Kazemimoghadam and Gu, 2021) use saliency maps for a specific use case (post-operative breast radiotherapy), (A. A. Ismail and Bravo, 2021) addresses that the existing saliency map generators result in noisy maps. Since the model gradients should highlight mostly only the relevant features for consistent and accurate performance, (A. A. Ismail and Bravo, 2021) proposed saliency-guided training that diminishes gradients on irrelevant features without worsening model performance. This is done through masking input features that results in lower gradient values, resulting in more sparse, precise gradients. (A. A. Ismail and Bravo, 2021) also provides experiments that incorporate the saliency-guided training on various modals (image, language, and time series) for classification tasks to show the framework's effectiveness.

(L. Wang and Metaxas, 2019) also focused on the general perspective on attention maps, where (L. Wang and Metaxas, 2019) recognized that there is substantial amount of overlaps between class-specific attention maps, which could lead to more "visual confusion" for models and classification errors. Hence, (L. Wang and Metaxas, 2019) established an end-to-end pipeline (called ICASC, Improving Classification with Attention Separation and Consistency) that provides the discrimination of class-specific attention and enforces the disciminative feature to be consistent across models' CNN layers to improve classification performance.

(K. Li and Fu, 2018) also developed an end-to-end network (called GAIN, Guided Attention Inference Network), which supervises a model's attention maps during training to guide the model to make predictions based on relevant features of the inputs. (K. Li and Fu, 2018) states that GAIN was developed to tackle the problem that attention maps (that are generated using only classification labels), which are used as priors for segmentation or localization tasks, tend to only cover smaller regions, and hence the maps are incomplete and less accurate. GAIN uses two network streams, one is for classification and the other is for attention mining, where the classification stream helps the other by providing information on areas of the inputs associated to classification task, while the other ensures that all relevant parts are incorporated during classification.

Overall, there is no one single architecture that outperforms others in terms of multi-task classification and segmentation especially for medical domain, but prior work shows that ensemble learning or multi-stage training performs better in general for the multi-tasking (E. Sogancioglu and Murphya, 2021). Additionally, much of the published works incorporate transfer learning done on large, general image datasets, but they are not much work done on medical datasets (Hwang and Kim, 2016). To the best of our knowledge, this research is the first to directly use heatmap generators and eye-gaze information in model training to guide the U-Net model to improve disease classification performance and to produce attention maps similar to how the radiologists view CXR images for diagnosis.

3. Methods

This section first gives an overview of the dataset (A. Karargyris and Moradi, 2021) used for this study and introduces two baseline experiments that are established in (A. Karargyris and Moradi, 2021), which showcased the usability of the dataset. One of the baseline experiments and its corresponding results are treated as the baseline model and baseline results for this study. The comparison and analysis of the baseline experiments' results are also detailed. Secondly, this section illustrates the motivation behind this research, which came from assessing the effectiveness of adding a segmentation component to a classifier. Finally, this section outlines this study's three proposed models, their architectures, and their training methods.

3.1 Dataset and Baseline Experiments

3.1.1 Dataset

To incorporate the heatmap generators and radiologist's eye-gaze data on CXR images when training deep learning models, this work used a dataset that contains 1083 chest radiographs that preserve the high image quality as DICOM files, which were reviewed and reported by one radiologist (A. Karargyris and Moradi, 2021). The dataset also contains the transcribed radiology report, the radiologist's dictation audio and eye gaze coordinates mapped onto the corresponding images, and the associated disease class labels. The class labels include "normal", "CHF", and "pneumonia". Several CXR image examples of each of the classes in the dataset can be seen in Figure 8. The labels were all from formal clinical diagnoses, and the dataset contains an equal number of datapoints for each class. Because there was a misalignment between the csv files provided in the dataset and several image IDs were missing when generating eye-gaze heatmaps, all the experiments ran in this work dealt with 1017 images from the dataset and their corresponding eye-gaze information. A list of the 1017 images used in this study is available on our github repository[1]. For training, this dataset was split into training, validation, and test sets with the percentage of 80, 10, and 10, respectively. When splitting, unique patient IDs were in only one of the training, validation, or test datasets to prevent potential biases.

---

[1] https://github.com/IMICSLab/Classification_and_Explainability_in_Radiology

The eye-tracking data was used to teach the model about how radiologists observed the CXR images, thus the eye-gaze coordinates and fixation data were utilized to generate temporal and static eye-gaze heatmaps (as can be seen in Figure 9) using an open-sourced code (A. Karargyris and Moradi, 2021) where static heatmaps are concatenations of the corresponding temporal heatmaps.

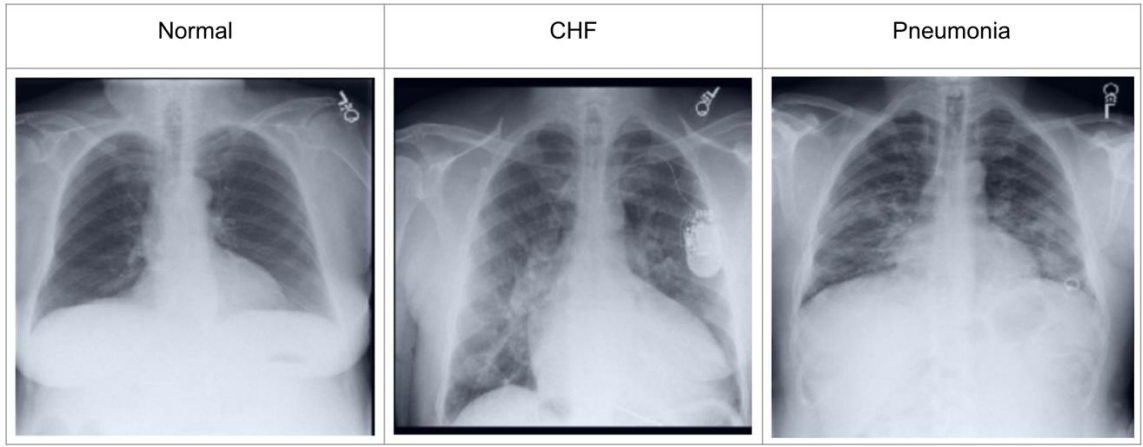

*Figure 8: Example CXR images for each of the three classes from (A. Karargyris and Moradi, 2021)*

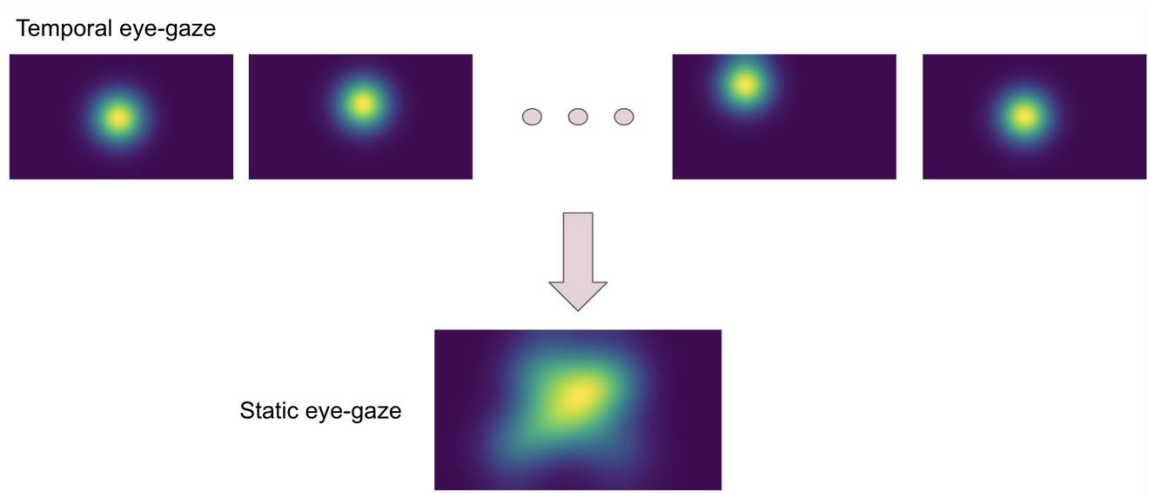

*Figure 9: Temporal and static eye-gaze heatmaps*

3.1.2 Baseline Experiments

The dataset that was used in this research was introduced in a paper (A. Karargyris and Moradi, 2021) that showed how the eye-gaze information could be utilized when training a deep learning model for a disease classification task using the area under the receiver operating characteristic curve (AUC) as the metric through conducting two experiments.

(A. Karargyris and Moradi, 2021)'s first set of experiments used both the DICOM CXR images and the temporal eye-gaze heatmaps as part of the inputs to the model as shown in Figure 10 to operate disease classification. This experiment was created to show how the temporal eye-gaze data in addition to the chest radiographs can be fed into a deep learning model as inputs.

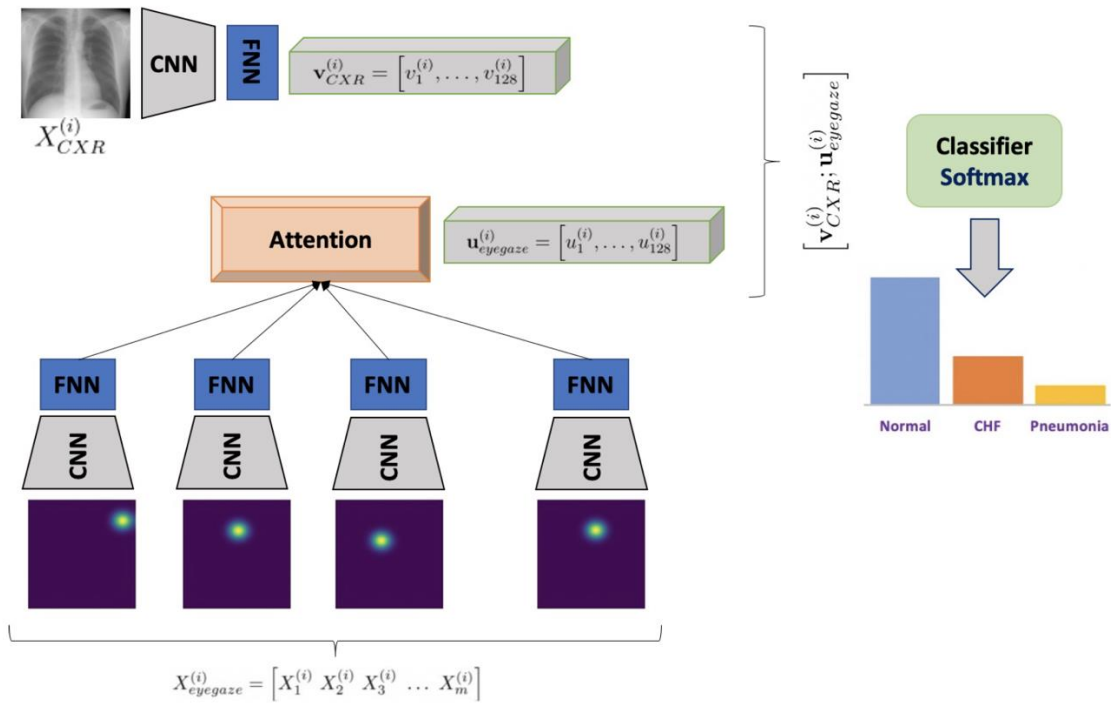

*Figure 10: (A. Karargyris and Moradi, 2021)'s first experiment's model architecture using the radiologist's temporal eye-gaze information to show how the eye-gaze heatmaps can be used in AI models*

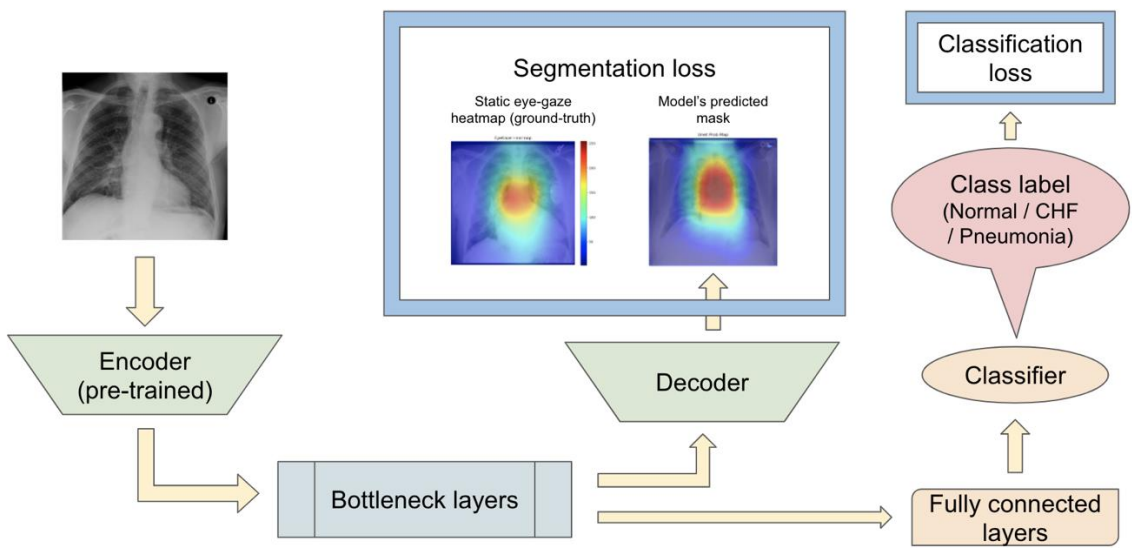

*Figure 11: (A. Karargyris and Moradi, 2021)'s second experiment's U-Net model architecture with the radiologist's static eye-gaze data used as the ground truths for the decoder's outputs*

(A. Karargyris and Moradi, 2021)'s second set of experiments was created to show how the eye-gaze data can be used for explainability purposes by using them in training so that the model can produce probability maps that look similar to the static eye-gaze heatmaps. It consisted of a U-Net structure as shown in Figure 11, with convolutional encoder and bottleneck layers that use pre-trained EfficientNet-b0 (Yakubovskiy, 2020; Tan and Le, 2020), a classification head, and a convolutional decoder. It computed and combined two sets of losses (both using Binary Cross Entropy with Logits Loss function); one was classification loss from the classification head, and the other was a segmentation loss computed between the static eye-gaze heatmap of the corresponding CXR image and the U-Net's output prediction mask from the decoder. The average AUC values achieved using this method are shown in Table 1. It appears that the model was able to classify the "normal" condition the best, while it struggled most to correctly classify the "pneumonia" condition.

*Table 1: AUC values for (A. Karargyris and Moradi, 2021)'s second experiment (the values in parenthesis are 2.5th and 97.5th percentile values)*

| Average AUC | "Normal" AUC | "CHF" AUC | "Pneumonia" AUC |
|---|---|---|---|
| 0.872 (0.840, 0.897) | 0.923 (0.895, 0.945) | 0.916 (0.871, 0.938) | 0.781 (0.713, 0.851) |

Some examples of the generated heatmaps using the trained U-Net model from (A. Karargyris and Moradi, 2021) can be seen in Figures 12, 13, and other corresponding outputs in the Appendix. The leftmost image is the model's input chest radiograph, the inner left heatmap is the output of running Grad-CAM after training, the inner right heatmap is the static eye-gaze heatmap from the dataset that was used as the ground truth when computing the segmentation loss, and the rightmost heatmap is the model's predicted probability mask generated from the decoder. Although the U-Net's predicted masks seemed to be trained well to align more with the static eye-gaze heatmaps as was expected from how the segmentation loss was computed, the Grad-CAM results for explainability purposes did not visually overlap well with the static eye-gaze heatmaps, and hence the model's focus when making classification predictions seemed to not necessarily be similar to the radiologist's eye-gaze focus when making diagnosis. This U-Net architecture, which used static eye-gaze heatmaps as part of the segmentation loss computation, and its corresponding AUC values were treated as the baseline model and baseline AUC values for this study.

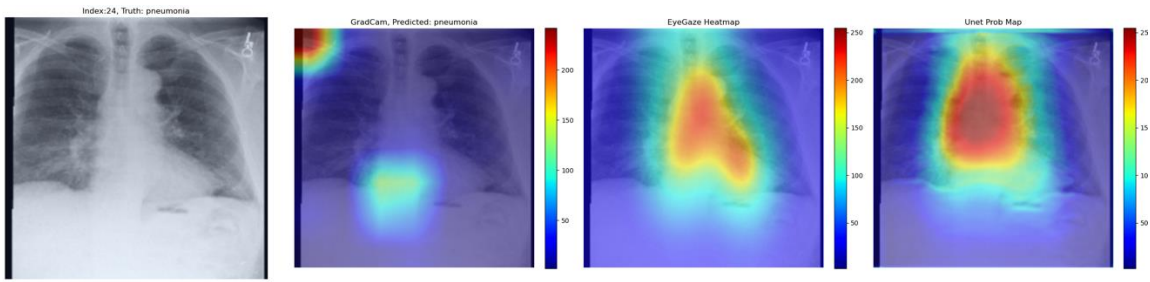

*Figure 12: Heatmap generated from Grad-CAM and the U-Net's decoder for pneumonia class correctly classified*

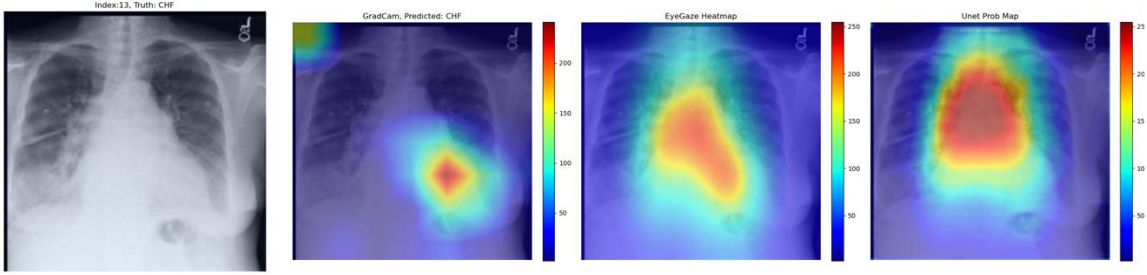

*Figure 13: Heatmap generated from Grad-CAM and the U-Net's decoder for CHF class correctly classified*

3.1.3 Motivation for this study

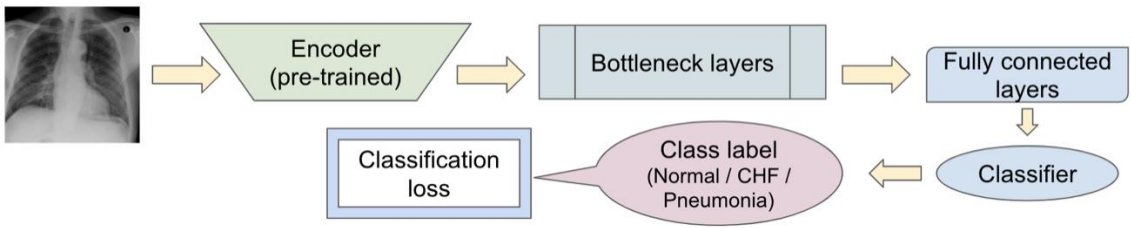

*Figure 14: Model architecture for the baseline model provided in the paper for the static eye-gaze experiment without the segmentation component*

*Table 2: Comparing the AUC values between (A. Karargyris and Moradi, 2021)'s baseline model with no segmentation component and the U-Net architecture (the values in parenthesis are 2.5th and 97.5th percentile values)*

| Model details | Average AUC [50th, (2.5th, 97.5th)] | "Normal" AUC | "CHF" AUC | "Pneumonia" AUC |
|---|---|---|---|---|
| U-Net (treated as the baseline model for this research) | 0.872 (0.840, 0.897) | 0.923 (0.895, 0.945) | 0.916 (0.871, 0.938) | 0.781 (0.713, 0.851) |
| Model without segmentation component | 0.873 (0.838, 0.908) | 0.878 (0.836, 0.918) | 0.934 (0.898, 0.975) | 0.805 (0.763, 0.886) |

To observe if adding the segmentation component to the disease classifier contributes to improving the overall classification AUC values, the performances of the U-Net architecture 11 and a model without the segmentation component 14 were compared using the classification AUC values. The model without the segmentation component was the baseline model that (A. Karargyris and Moradi, 2021) provided for the static eye-gaze heatmap experiment. This model did not attempt to enhance the explainability component simultaneously and only focused on the classification task. The classification performance of the non-segmentation model and the U-Net using the AUC metric, which is depicted in Table 2, shows that the overall average AUC values are similar for both models. From these AUC results, it appears that the multitasking of improving classification and incorporating the explainability component to model training did not result in improved mean AUC values. Nevertheless, when the AUC values for each of the three classes were viewed separately, it was concluded that U-Net architecture improved to classify "normal" condition better than the non-segmentation model, but "CHF" and "pneumonia" classes had lower AUC values compared to those of the non-segmentation model. Such observations suggests that there could be improvements made for this U-Net architecture that would incorporate the explainability component when attempting to improve the overall classification AUC, particularly for the "CHF" and "pneumonia" classes.

3.2 Defining segmentation loss using heatmap generators

Given the different heatmap generators for explainability improvement, (A. Karargyris and Moradi, 2021)'s two experiments, and the dataset, this study produced and experimented with three sets of proposed models using the (A. Karargyris and Moradi, 2021)'s U-Net architecture to observe if using both the eye-gaze data and various heatmap generators during model's training time could result in improved AUC for the classification task while enhancing the predicted mask generation. Specifically, the experiments involved using the heatmap generators (such as guided back-propagation and deconvNet) during training and different ways of using the heatmaps when computing the segmentation loss.

Since the major goal was to improve the classification AUC, the average AUC value and the AUC values for each of the three classes were used to evaluate the models' performance on each of the experiments. Finally, the 95% confidence interval (CI) (with 2.5th % and 97.5th % values) in addition to the average AUC values were measured for each of the three class labels over 55 to 60 samples with resampling for 30 iterations using the test set.

3.2.1 Proposed Model Set 1

For the first proposed model set, the segmentation loss was computed using heatmaps generated from the different generators (which included guided back-propagation and deconvNet) during the training time against the static eye-gaze heatmaps, which were treated as ground truth as can be seen in Figure 15. This experiment attempted to guide the model's trainable parameters to mimic the static eye-gaze heatmaps so that the model could learn to focus on similar areas of the chest radiographs as how the radiologist did.

3.2.2 Proposed Model Set 2

The second proposed model set was run with another modification to the network training when computing the segmentation loss. In contrast to the previous proposed model that computed the segmentation loss using the heatmaps generated from either guided back-propagation or deconvNet and the static eye-gaze heatmaps, the segmentation loss for this proposed model was computed using the differences between the model's predicted masks (which are the outputs of the decoder) and heatmaps generated from either guided back-propagation or deconvNets as can be seen in Figure 16. This experiment was run to assess if the model can be trained to produce predicted probability masks to appear consistent to the heatmap generator's outputs.

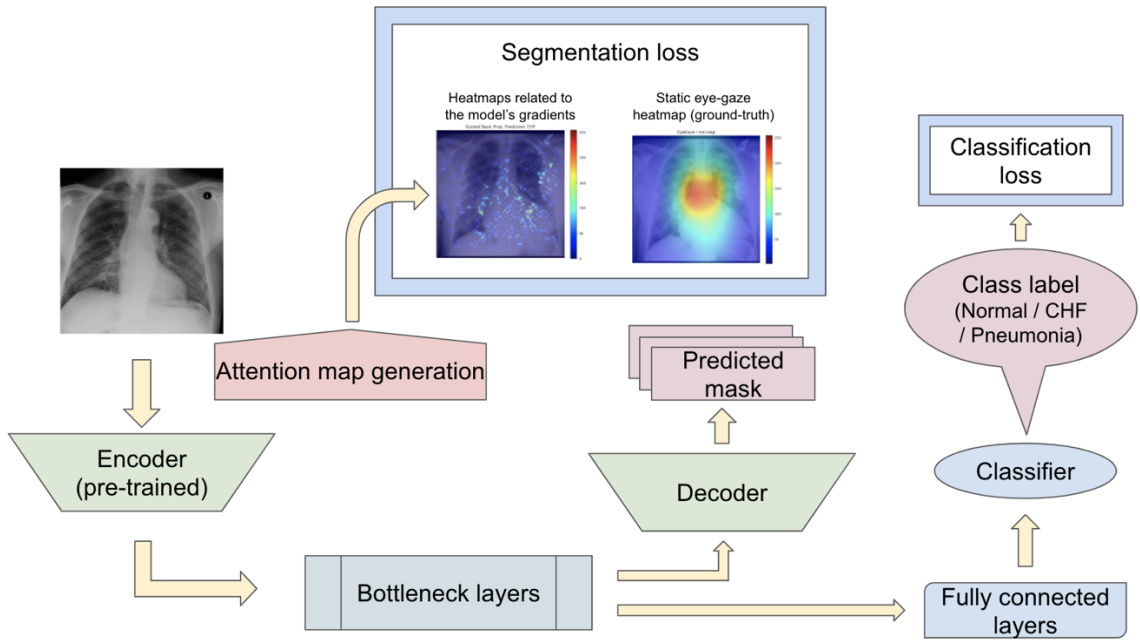

*Figure 15: Model architecture for the proposed model set 1 with a unique segmentation loss computation*

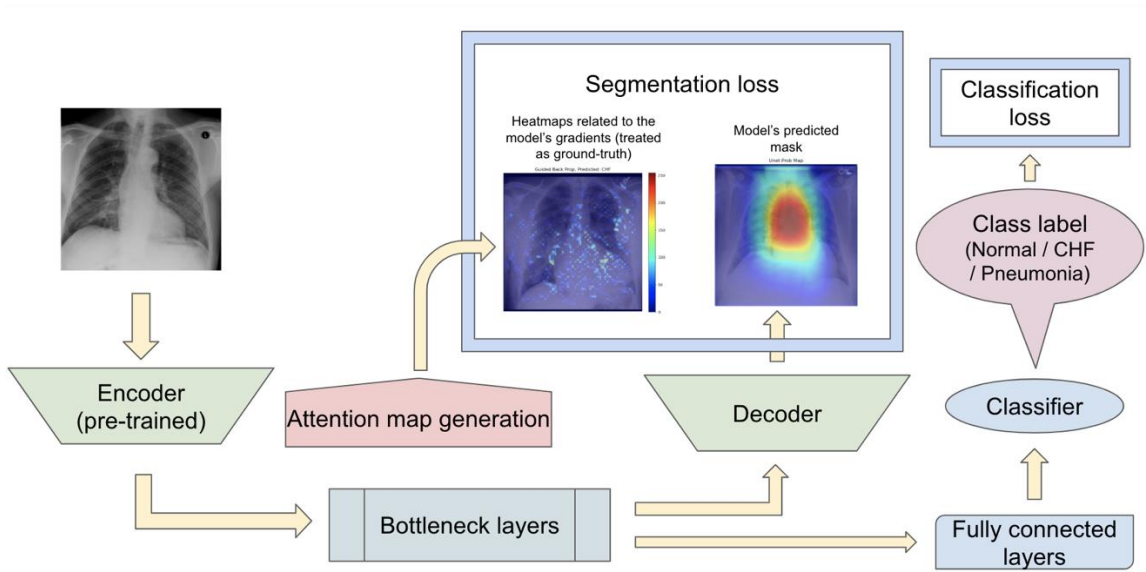

*Figure 16: Model architecture for the proposed model set 2 with a unique segmentation loss computation*

3.2.3 Proposed Model Set 3

Since the previous two proposed model sets did not focus on training the model's decoder to generate predicted masks that appear similar to the static eye-gaze heatmaps, the third proposed model set used a combination of two segmentation losses, where one of which was computed using the differences between the U-Net's predicted masks and the static eye-gaze heatmaps, while the other loss was computed using the differences between the heatmaps generated from one of the generators and the static eye-gaze heatmaps as shown in Figure 17. This modification in segmentation loss computation incorporated the different heatmap generators so that the model's trainable parameters would learn to focus on similar areas as the static eye-gaze heatmaps, while guiding the model's decoder to produce predicted masks that look similar to the static eye-gaze heatmaps. The ratio of the combination of the two segmentation losses was treated as an adjustable hyperparameter.

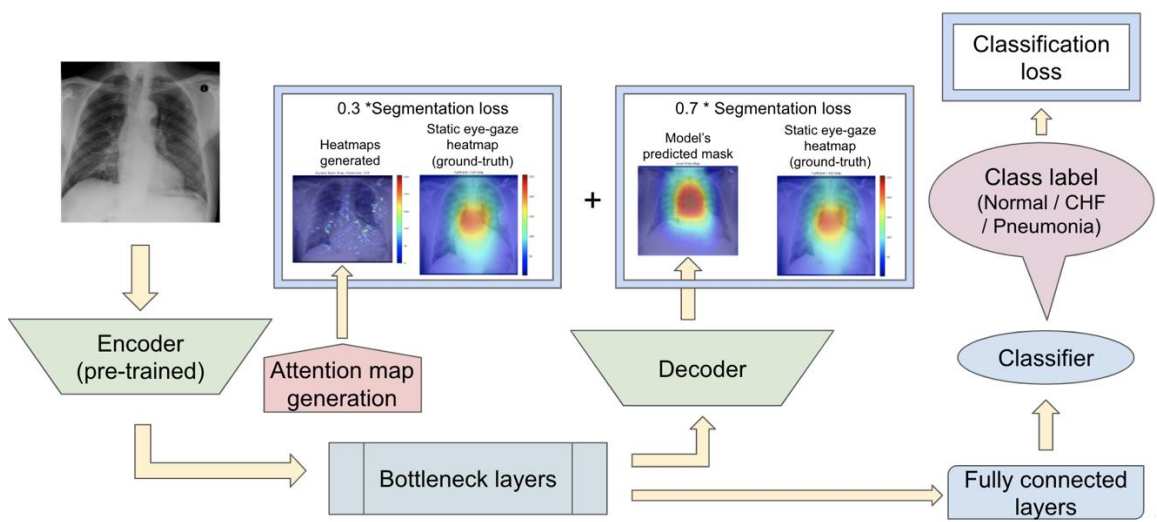

*Figure 17: Model architecture for the proposed model set 3 with a unique segmentation loss computation*

4. Results

4.1 Proposed Model Set 1

For the first experiment set, the use of guided back propagation as the heatmap generator when computing the segmentation loss resulted in the highest average AUC values compared to the AUC values obtained using deconvNet and the baseline AUC values as can be seen in Table 3. This difference between the use of guided back-propagation and deconvNets is understandable due to the nature of the computation differences between guided back-propagation and deconvNets, where guided back-propagation is based on deconvNet but sets the negative gradients for inputs to zero, and hence guided back-propagation highlights the important regions of inputs even more (Y. Lianga and Jiang, 2021). For both guided back-propagation and deconvNet cases, the average AUC values were higher than that of the baseline model, and the greatest improvement in AUC values occurred for classifying "pneumonia" with 5% increase when using guided back-propagation and 2.5% increase when using deconvNet. Additionally, classifying "CHF" also had 3% improvement when using guided back-propagation and 1.9% improvement when using deconvNet. Although both "CHF" and "pneumonia" classification improved in this experiment set with the use of the heatmap generators during the model's training time, this experiment did not result in a great improvement for average AUC values because the AUC values for classifying the "normal" class was lower with the use of the heatmap generators when computing segmentation loss compared to the baseline values.

*Table 3: Comparing the output AUC values between (A. Karargyris and Moradi, 2021)'s U-Net model and a U-Net with different heatmap generators for the Experiment Set 1 (the values in parenthesis are 2.5th and 97.5th percentile values)*

| Model and heatmap generator's details | Average AUC [50th, (2.5th, 97.5th)] | "Normal" AUC | "CHF" AUC | "Pneumonia" |
|---|---|---|---|---|
| U-Net (baseline) | 0.872 (0.840, 0.897) | 0.923 (0.895, 0.945) | 0.916 (0.871, 0.938) | 0.781 (0.713, 0.851) |
| Guided back-propagation | 0.891 (0.847, 0.939) | 0.896 (0.854, 0.945) | 0.946 (0.891, 0.979) | 0.831 (0.763, 0.923) |
| DeconvNet | 0.884 (0.843, 0.935) | 0.916 (0.884, 0.943) | 0.935 (0.905, 0.959) | 0.806 (0.715, 0.918) |

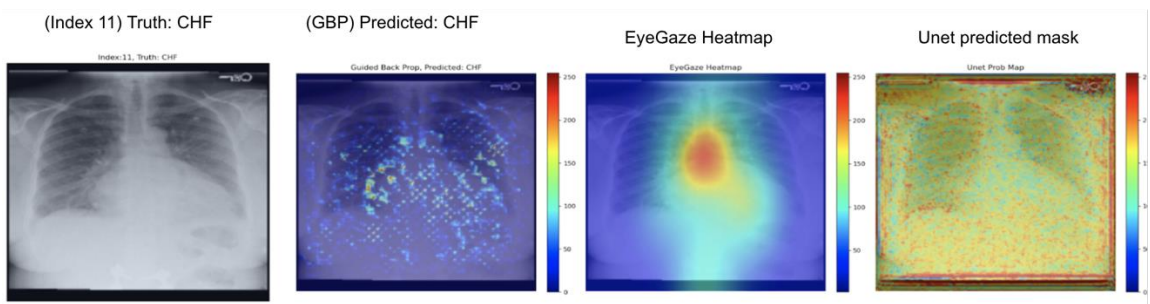

*Figure 18: Model's output heatmaps using guided back-propagation as the heatmap generator for Experiment Set 1 (for the correctly classified CHF class)*

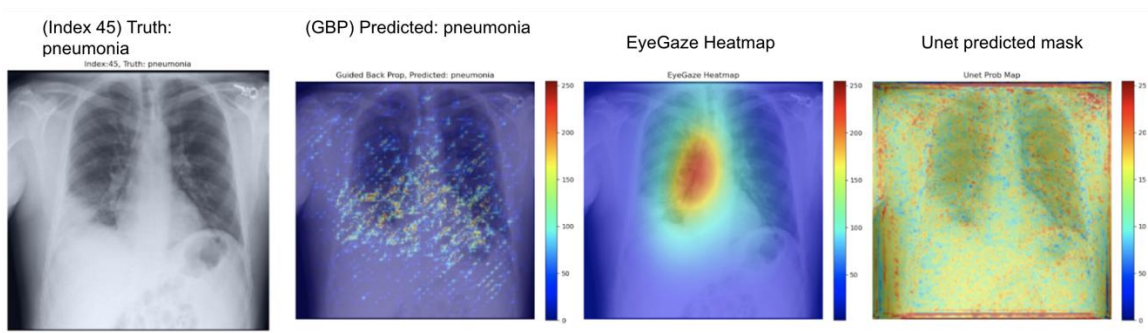

*Figure 19: Model's other output heatmaps using guided back-propagation as the heatmap generator for Experiment Set 1 (for the correctly classified pneumonia class)*

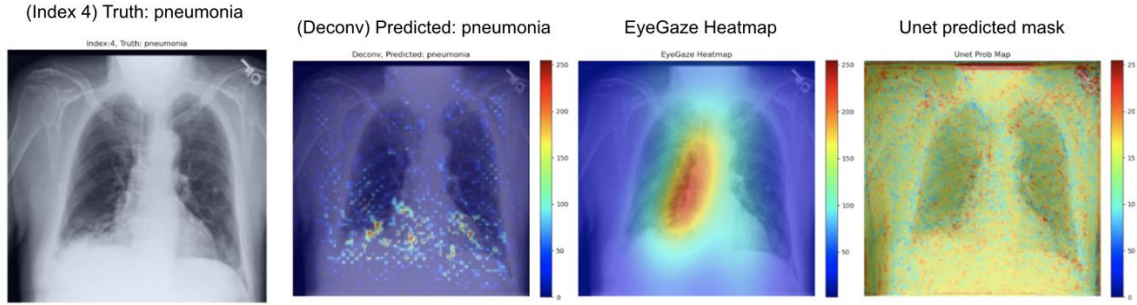

*Figure 20: Model's output heatmaps using deconvNet as the heatmap generator for Experiment Set 1 (for the correctly classified pneumonia class)*

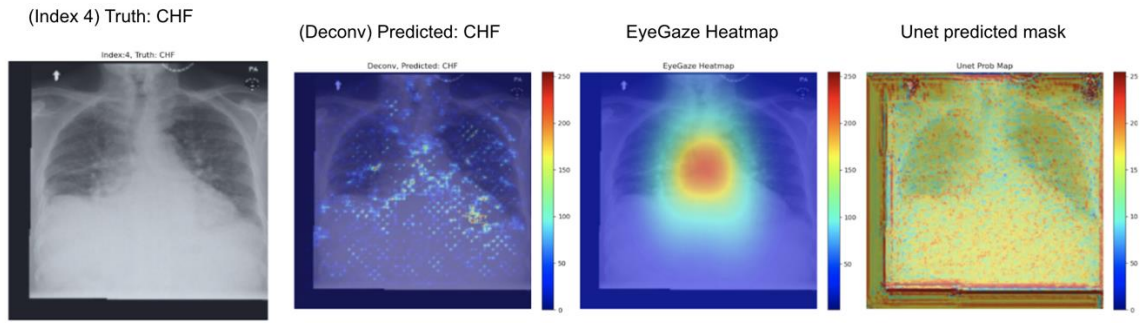

*Figure 21: Model's other output heatmaps using deconvNet as the heatmap generator for Experiment Set 1 (for the correctly classified CHF class)*

As can be seen in Figures 18, 19, 20, 21, and other outputs in the Appendix, the guided back-propagation and deconvNet's heatmaps highlighted areas near the lungs and heart similarly to the static eye-gaze heatmaps. Additionally, there was a correlation between the greater vividness of guided back-propagation's heatmaps and the slightly higher AUC values compared to those of deconvNets as can be seen in Figure 3. However, the model's probability masks appeared to be quite different from that of static eye-gaze heatmaps. This was because the segmentation loss was computed without the U-Net's prediction mask in attempts to align the model's parameters to appear similar to the static eye-gaze heatmaps, and hence the decoder was not trained to produce prediction masks that look similar to the static eye-gaze heatmaps.

4.2 Proposed Model Set 2

Similarly to the first proposed model set, the guided back-propagation method performed the best compared to the baseline and the deconvNet method. Furthermore, the outcome of using guided back-propagation for this proposed model set had a slight improvement in the average AUC value compared to the first proposed model set. The greatest improvement occurred for the "pneumonia" class AUC, with 6.9% increase in AUC, which contributed to the 2.5% increase in the average AUC values for using guided back-propagation compared to the baseline AUC values.

*Table 4: Comparing the output AUC values between (A. Karargyris and Moradi, 2021)'s U-Net model and a U-Net with different heatmap generators for the Experiment Set 2 (the values in parenthesis are 2.5th and 97.5th percentile values)*

| Model and heatmap generator's details | Avg AUC [50th, (2.5th, 97.5th)] | "Normal" AUC | "CHF" AUC | "Pneumonia" AUC |
|---|---|---|---|---|
| U-Net (baseline) | 0.872 (0.840, 0.897) | 0.923 (0.895, 0.945) | 0.916 (0.871, 0.938) | 0.781 (0.713, 0.851) |
| Guided back-propagation | 0.897 (0.848, 0.921) | 0.907 (0.863, 0.946) | 0.922 (0.888, 0.956) | 0.850 (0.792, 0.900) |
| DeconvNet | 0.881 (0.828, 0.907) | 0.885 (0.836, 0.928) | 0.922 (0.890, 0.952) | 0.827 (0.732, 0.901) |

For this experiment, with the use of deconvNet as the heatmap generator, the U-Net was trained so that the decoder would attempt to mimic the heatmaps generated from the generators when outputting its prediction mask as can be seen in rightmost images of Figures 22 and 23.

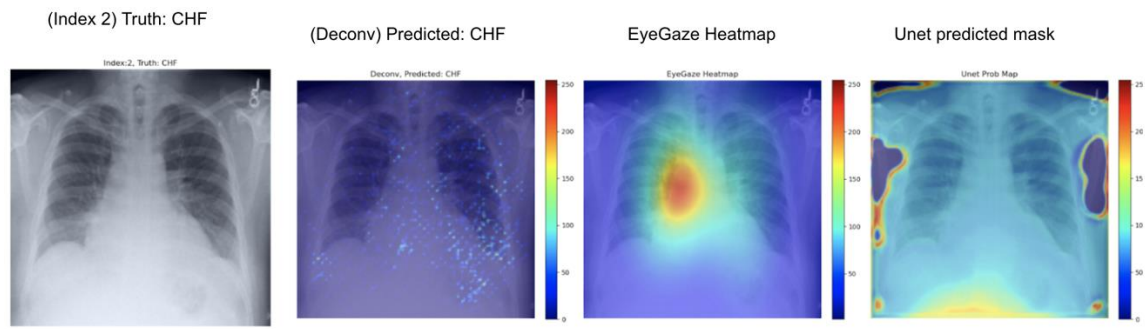

*Figure 22: Model's output heatmaps using deconvNet as the heatmap generator for Experiment Set 2 (for the correctly classified CHF class)*

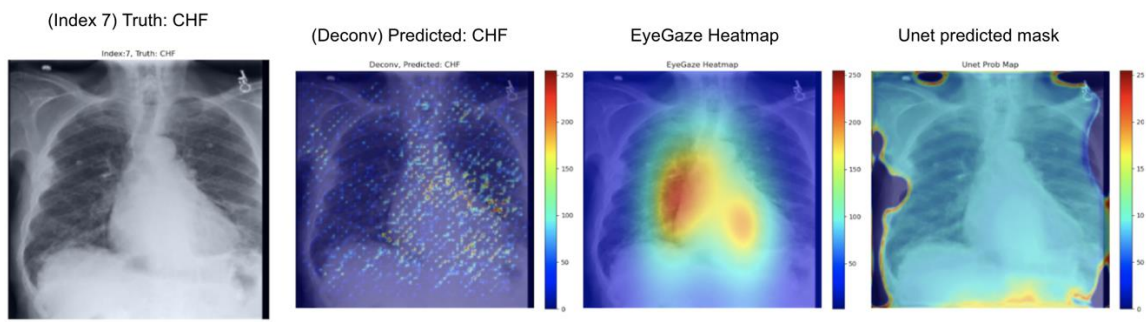

*Figure 23: Model's output heatmaps using deconvNet as the heatmap generator for Experiment Set 2 (for the correctly classified CHF class)*

The use of guided back-propagation as the heatmap generator (outputs shown in Figures 24 and 25) resulted in the U-Net's generated predicted masks having higher intensity than those with deconvNets. More outputs of the model can be viewed in the Appendix.

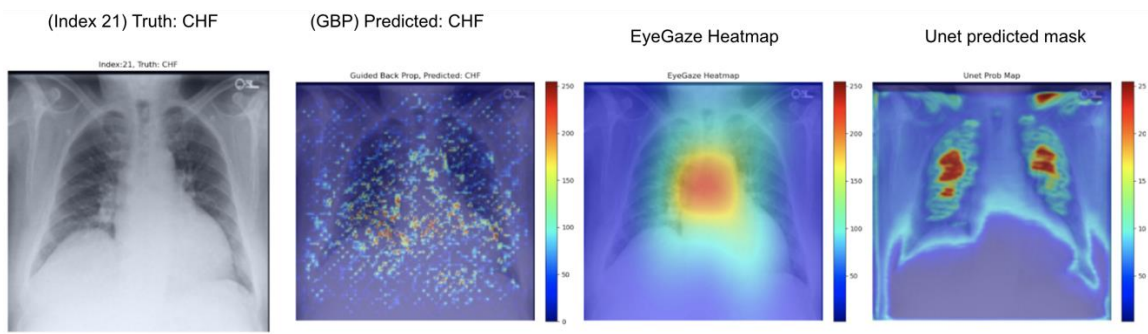

*Figure 24: Model's output heatmaps using guided back-propagation as the heatmap generator for Experiment Set 2 (for the correctly classified CHF class)*

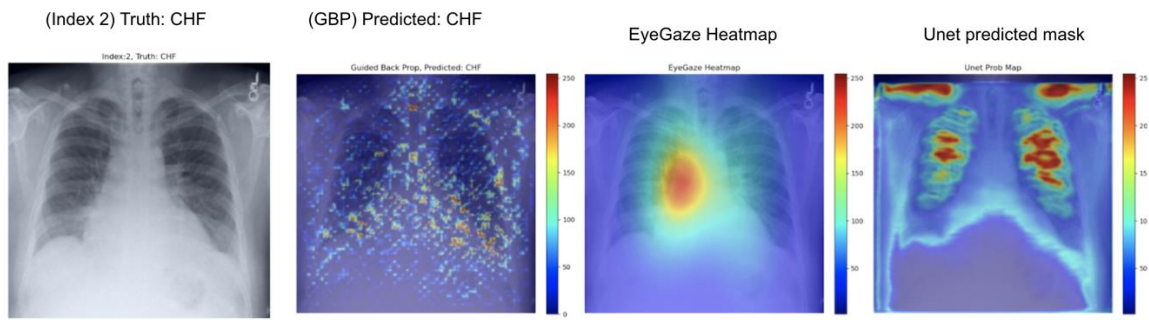

*Figure 25: Model's other output heatmaps using guided back-propagation as the heatmap generator for Experiment Set 2 (for the correctly classified CHF class)*

Although the U-Net's probability masks did not mimic the generated heatmaps, there seems to be a correlation between the heatmaps' intensity differences and the differences in the U-Net's predicted masks' intensity and in the locations of focus when using different heatmap generators. Additionally, there was a correlation between the greater intensity of the heatmaps and the predicted masks and the higher average AUC values when comparing guided back-propagation and deconvNet, which can be seen in Figure 4.

4.3 Proposed Model Set 3

This experiment with the combination of two segmentation losses incorporated the heatmap generators in training time to guide the model's parameters to focus on similar areas as the radiologist viewed, while guiding the model to produce predicted masks that look similar to the static eye-gaze heatmaps for explainability purposes.

As can be seen in Table 5, the use of deconvNet resulted in lower AUC values than the baseline model except for the "CHF" classification, which improved by 1.8%. These AUC results appear to be associated with how the U-Net's decoder was unable to mimic the static eye-gaze heatmaps when creating the predicted probability mask as can be seen in Figures 26 and 27.

*Table 5: Comparing the output AUC values between (A. Karargyris and Moradi, 2021)'s U-Net model and a U-Net with different heatmap generators for the Experiment Set 3 (the values in parenthesis are 2.5th and 97.5th percentile values)*

| Model and heatmap generator's details | Avg AUC [50th, (2.5th, 97.5th)] | "Normal" AUC | "CHF" AUC | "Pneumonia" AUC |
|---|---|---|---|---|
| U-Net (baseline) | 0.872 (0.840, 0.897) | 0.923 (0.895, 0.945) | 0.916 (0.871, 0.938) | 0.781 (0.713, 0.851) |
| Guided back-propagation and predicted mask | 0.913 (0.860, 0.966) | 0.921 (0.866, 0.968) | 0.962 (0.933, 0.989) | 0.859 (0.732, 0.957) |
| DeconvNet and predicted mask | 0.860 (0.806, 0.907) | 0.907 (0.856, 0.962) | 0.934 (0.891, 0.974) | 0.741 (0.643, 0.840) |

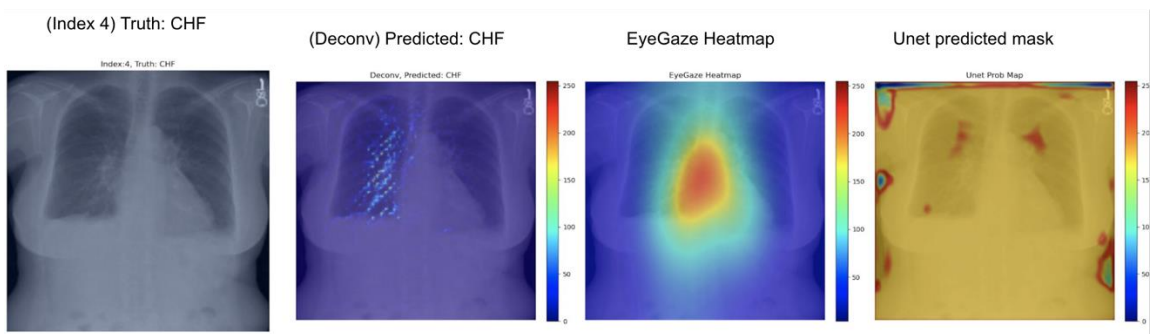

*Figure 26: Model's output heatmaps using deconvNet as the heatmap generator for Experiment Set 3 (for the correctly classified CHF class)*

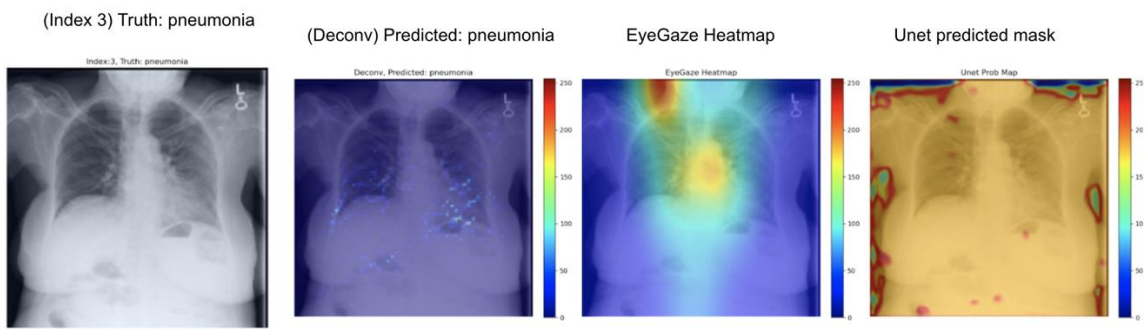

*Figure 27: Model's output heatmaps using deconvNet as the heatmap generator for Experiment Set 2 (for the correctly classified pneumonia class)*

On the contrary, the use of guided back-propagation for this segmentation loss computation had improvements in most of the AUC values, and the model's decoder was also able to generate the predicted mask that highlighted similar areas as the static eye-gaze heatmaps as shown in Figures 28, 29, and other corresponding outputs in the Appendix. Specifically, when using guided back-propagation, the average AUC value improved by 4.1%, the "CHF" classification improved by 4.6%, and the "pneumonia" classification improved by 7.8% compared to the baseline model that did not use the heatmap generators during the training time. Hence, this modification in the segmentation loss computation with guided back-propagation produced the best AUC values across all of the experiments in this research.

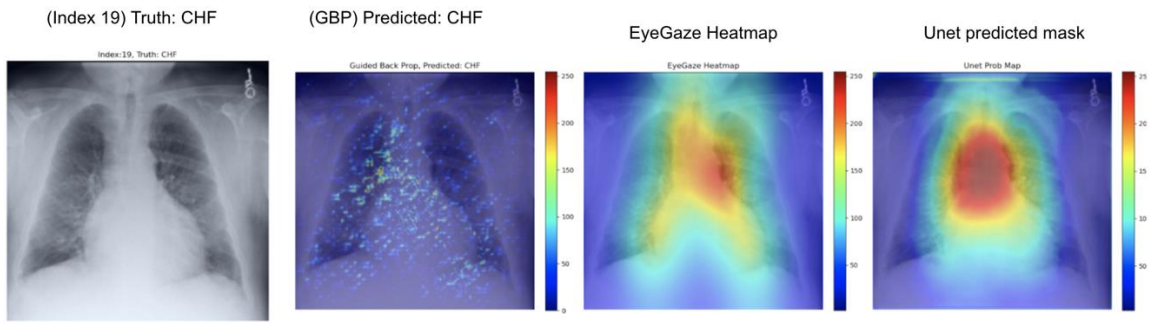

*Figure 28: Model's output heatmaps using guided back-propagation as the heatmap generator for Experiment Set 3 (for the correctly classified CHF class)*

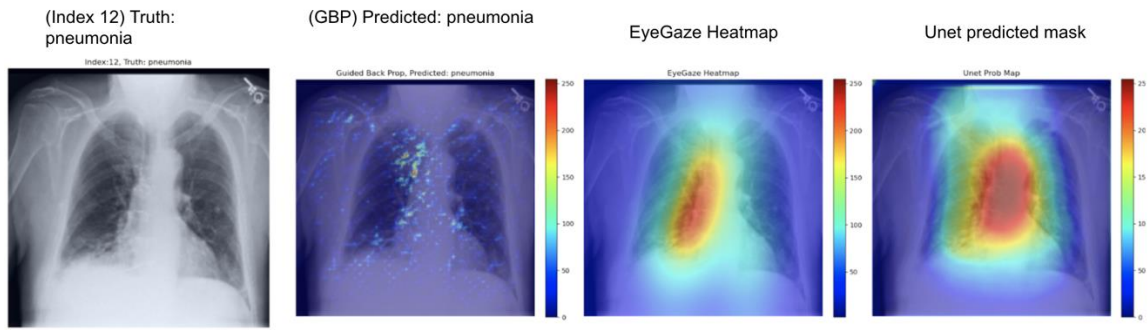

*Figure 29: Model's output heatmaps using guided back-propagation as the heatmap generator for Experiment Set 3 (for the correctly classified pneumonia class)*

Overall, the vividness of the heatmaps generated and the similarity between the U-Net's predicted masks and the static eye-gaze heatmaps seemed to be correlated with how well the average AUC values improved for the experiments. This suggested that the radiologist's eye-gaze information was valuable in improving the classification performance when used with the heatmap generators during model training.

5. Discussion

AI models for radiology are improving dramatically for disease classification or localization tasks, but there is often a trade-off between performance and explainability, as higher performing models tend to be deeper and more complex. Hence, explainability for AI in medical domain is gaining more attention to give insight into how the models arrived at their predictions and to increase trust in the use of such models. Additionally, from the legal point of view, although explainability is currently not a strict requirement for the AI use in clinical situations, FDA states that some level of transparency, which can include inputs and outputs of the AI model and its algorithm, is required to ensure transparency to the patients and the physicians (J. Amann and Madai, 2020). Furthermore, explainaibility will likely become a stricter requirement as more AI models are incorporated to clinical use (P. Hacker and Naumann, 2020). Therefore, this study highlights incorporating the explainability aspect into both model training and model outputs with the goal of improving disease classification.

There exist several AI studies that together classify diseases and generate heatmaps. Nonetheless, they have several issues, such as the publicly available datasets that such research uses sometimes contain incorrect labels obtained from NLP and the research oftentimes use only the CXR images and class labels without incorporating the other methods or data the radiologists usually use when making diagnoses.

Thus, this research proposed three model sets that take a U-Net architecture with distinct loss function computations using a dataset that contains eye-gaze information from a radiologist. This study showed that incorporating eye-gaze data and heatmap generators in model training can improve disease classification AUC, especially for "CHF" and "pneumonia" classes that help reduce false negatives. Moreover, by guiding the model's gradients to align to the radiologist's eye-gaze heatmaps, the model is able to produce enhanced heatmaps that can be used for explainability. Furthermore, this study confirms that using different types of data other than just the CXR images and disease class labels and following radiologists' methods for diagnosis when establishing and training AI models can increase the classification performance.

The heatmap visualisation outputs of the three proposed model sets highlight areas of CXR images that the model focused on when making predictions. The outputs of the first proposed model set for both guided back-propagation and deconvNet uses show that each generator appeared to focus on similar areas as the eye-gaze heatmaps did, but the U-Net's predicted masks seemed to not focus well, which is understandable due to the way segmentation loss was calculated for this model set.

The outputs of the second proposed model set using deconvNet as the heatmap generator (as can be seen in Figures 22 and 23) show that since the heatmap generators created more sparse intensities compared to the static eye-gaze heatmaps, the intensity of the model's predicted masks was also lower and covered a wider area of the chest compared to the outputs from the other experiments as can be seen on the right most images of Figures 22 and 23. On the contrary, there seemed to be a correlation between greater vividness and intensity for U-Net's predicted mask and the higher AUC values when comparing between the use of different generators.

Finally, the visualization outputs of the third proposed model set clearly showcased that the use of guided back-propagation method as the heatmap generator for this training method performed superior compared to others. Both the U-Net's predicted masks and the heatmaps generated from the generators focused on similar areas as the eye-gaze heatmaps, suggesting that the model is learning to generate explainable heatmaps while predicting the disease class.

Viewing the results from the three experiment sets, particularly focusing on the AUC values in Figure 5, the multitasking of improving classification and generating reasonable heatmaps for enhancing explainability during model training using the U-Net architecture was shown to be effective when applying guided back-propagation as the heatmap generator. Although the confidence intervals of the AUC values for the Proposed Model Set 3's guided back-propagation method had some overlap with the baseline's confidence intervals, this new method not only had improvements in the average AUC value (4.1%), but it also had greater improvements for the "CHF" (4.6%) and "penumonia" (7.8%) classes, which were the classes the baseline struggled to classify. Since the major focus for disease classification tasks is on correctly identifying the non-"normal" disease classes and decreasing the false negatives, these improvements with the use of guided back-propagation heatmap generator during the training time was significant. Furthermore, the experiments' results suggest that the use of heatmap generators (particularly guided back-propagation) in training time could also enhance the model's predicted masks generation for explainability purposes to better convey where the model's attention was on the input chest radiographs.

The major limitation of this research is the small size of the dataset and the potential bias the dataset may have because the eye-gaze data was collected from and the diagnosis was done by one radiologist (A. Karargyris and Moradi, 2021). Thus, other medical imaging datasets that are larger and contain eye-gaze information may be used to further validate the disease classification improvements produced by using heatmap generators and eye-gaze data during model training.

To further validate the classification decisions this study's proposed models make and to increase more trust in the model's predictions, DOCTOR's Totally Black Box scenario (F. Granese and Piantanida, 2021) can be used after the model makes class predictions on given CXR images. DOCTOR is a discriminator that can be used to detect if each of the predictions made by the AI model can be trusted or not irrespective of in- or out- distribution the data is coming from, and it does not require any prior knowledge on the dataset or model architecture. DOCTOR's decisions can signal physicians to take a second look at the CXR images and the associated class prediction the model made if DOCTOR rejects the predictions. When applying DOCTOR, there will be a trade-off between the number of samples that are rejected and the threshold value for false detection and acceptance rate, and ensuring that more misclassified samples are being rejected results in more samples including some that are truly correctly classified being rejected as well. Nonetheless, employing DOCTOR in this research's proposed model will be valuable because it is essential to eliminate model's misclassifications particularly in radiology field and the overall medical domain.

6. Conclusion

In this research, three proposed model sets that use heatmap generators in the U-Net model's training time were investigated for the radiology field to simultaneously improve the disease classification performance and better highlight the model's attention spots by generating explainable heatmaps. The experiments used a dataset that contain chest radiographs, eye-gaze coordinates from one radiologist, and the corresponding images' class labels. The confidence intervals of the AUC values for each experiment that incorporated the static eye-gaze information and the heatmap generators during the model's training time were computed. The proposed model in Figure 17 that used the weighted average of two segmentation losses (one computed between the heatmaps generated from guided back-propagation and static eye-gaze heatmaps and the other computed between the predicted masks from the U-Net's decoder and the static eye-gaze heatmaps) performed superior compared to the baseline model provided in (A. Karargyris and Moradi, 2021) and the other models that were tested in this research. As a future work, a larger medical imaging dataset with eye-gaze information from multiple radiologists could be used to further validate the disease classification improvement achieved by this method. Moreover, DOCTOR can be employed to estimate how much each of the predictions made by the model is trustworthy or not to further increase trust in the model's predictions.

Appendix A. Outputs of (A. Karargyris and Moradi, 2021)'s U-Net (Treated as the baseline for this research)

Figures 30, 31, and 32 were the outputs of the (A. Karargyris and Moradi, 2021)'s U-Net model (the baseline model for this research).

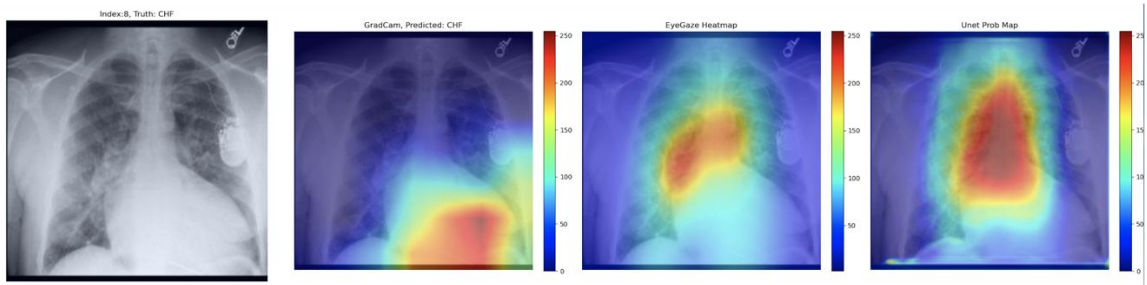

*Figure 30: Output of the paper's U-Net model with Grad-CAM ("CHF" correctly classified)*

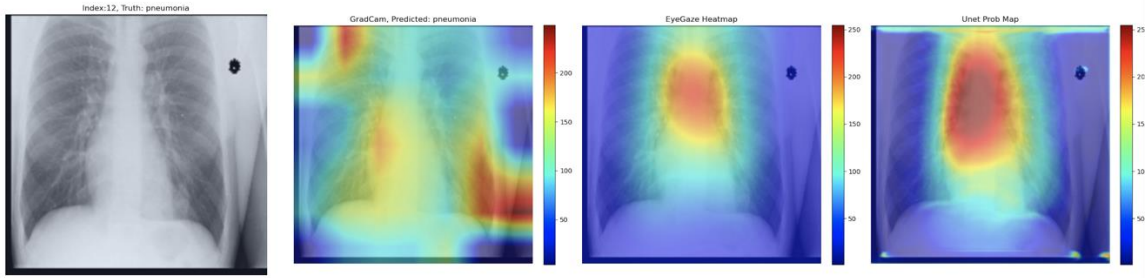

*Figure 31: Output of the paper's U-Net model with Grad-CAM ("pneumonia" correctly classified)*

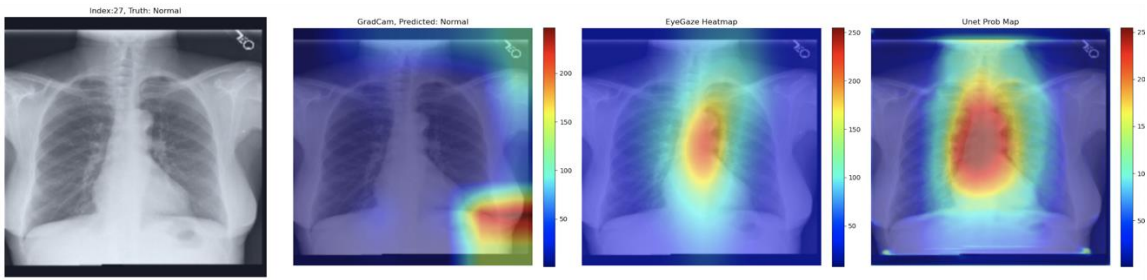

*Figure 32: Output of the paper's U-Net model with Grad-CAM ("normal" correctly classified)*

Appendix B. Outputs of Experiment Set 1

Appendix B.1 Outputs with guided back-propagation

Figures 33, 34, and 35 were the outputs of the Experiment Set 1 using guided back-propagation as the heatmap generator.

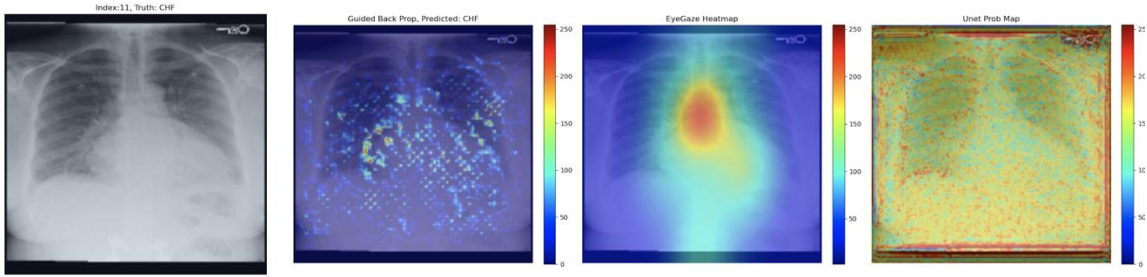

*Figure 33: Model's output heatmaps using guided back-propagation as the heatmap generator for Experiment Set 1 (for "CHF" class correctly classified)*

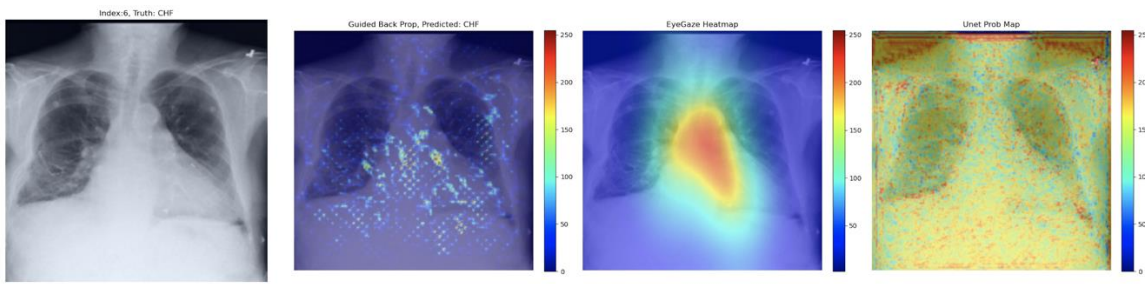

*Figure 34: Model's output heatmaps using guided back-propagation as the heatmap generator for Experiment Set 1 (for "CHF" class correctly classified)*

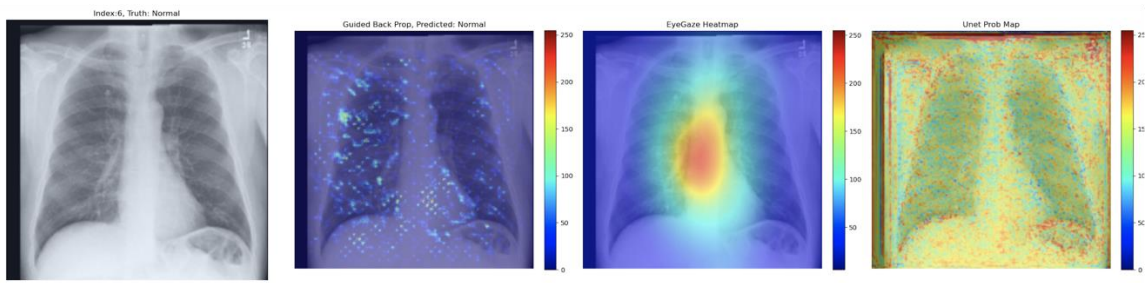

*Figure 35: Model's output heatmaps using guided back-propagation as the heatmap generator for Experiment Set 1 (for "normal" class correctly classified)*

Appendix B.2 Outputs with deconvNet

Figures 36, 37, and 38 were the outputs of the Experiment Set 1 using deconvNet as the heatmap generator.

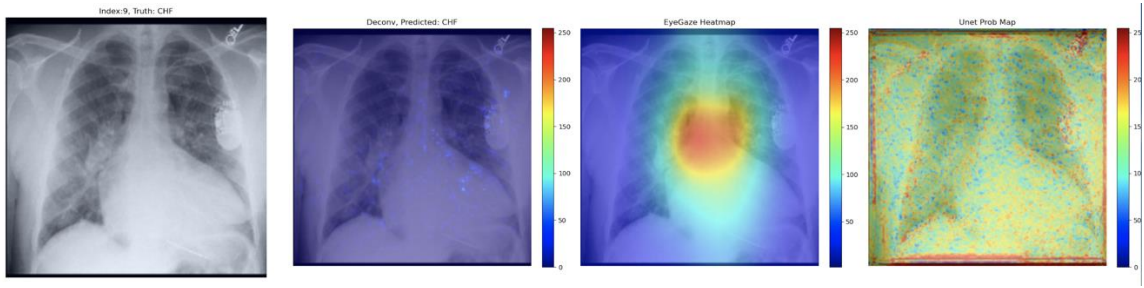

*Figure 36: Model's output heatmaps using deconvNet as the heatmap generator for Experiment Set 1 (for "CHF" class correctly classified)*

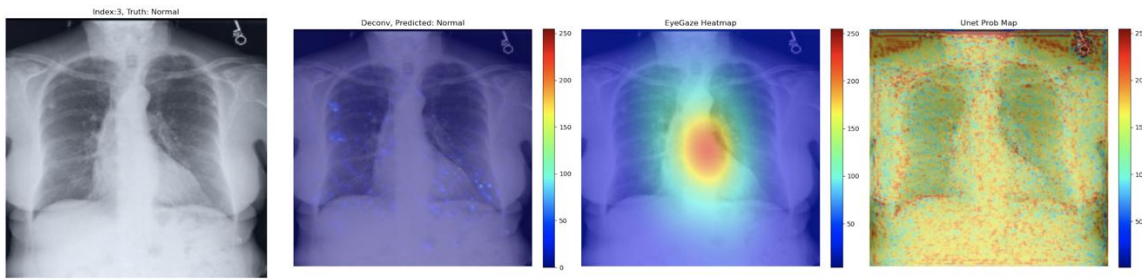

*Figure 37: Model's output heatmaps using deconvNet as the heatmap generator for Experiment Set 1 (for "normal" class correctly classified)*

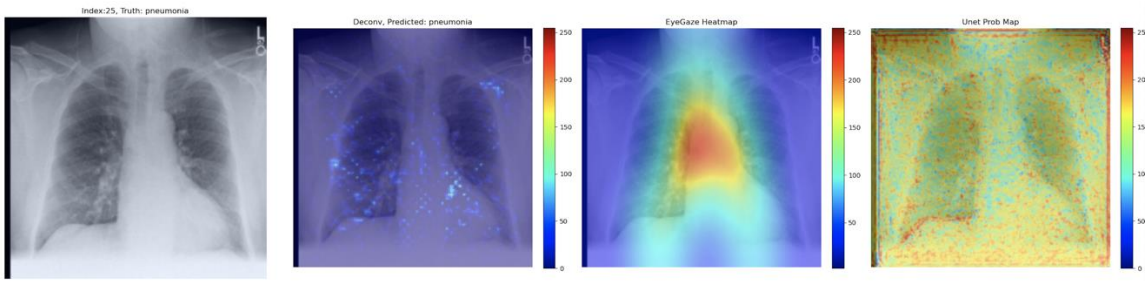

*Figure 38: Model's output heatmaps using deconvNet as the heatmap generator for Experiment Set 1 (for "pneumonia" class correctly classified)*

Appendix C. Outputs of Experiment Set 2

Appendix C.1 Outputs with guided back-propagation

Figures 39, 40, and 41 were the outputs of the Experiment Set 2 using guided back-propagation as the heatmap generator.

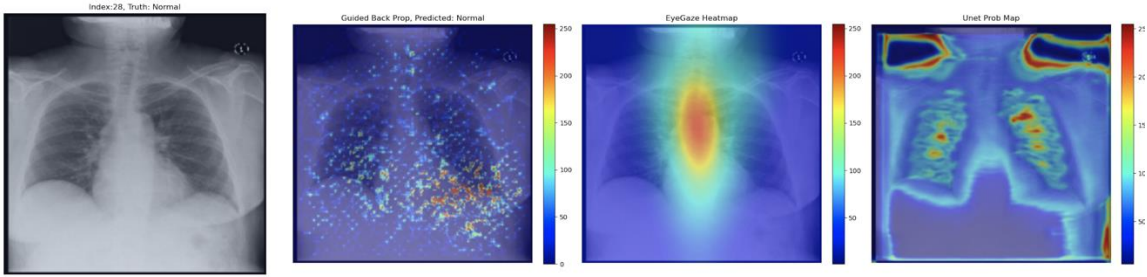

*Figure 39: Model's output heatmaps using guided back-propagation as the heatmap generator for Experiment Set 2 (for "normal" class correctly classified)*

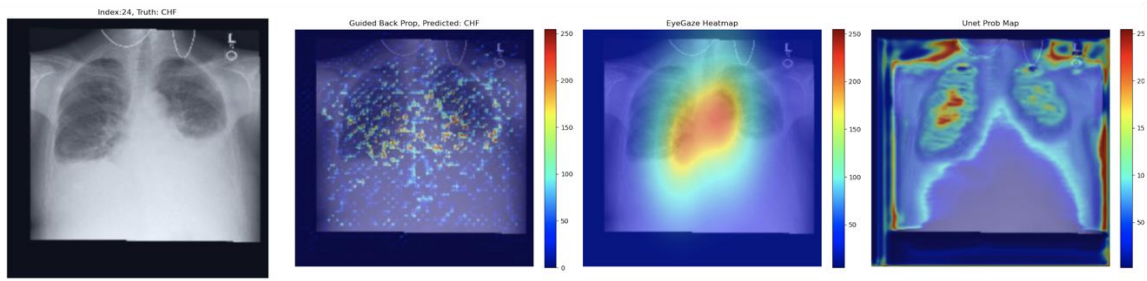

*Figure 40: Model's output heatmaps using guided back-propagation as the heatmap generator for Experiment Set 2 (for "CHF" class correctly classified)*

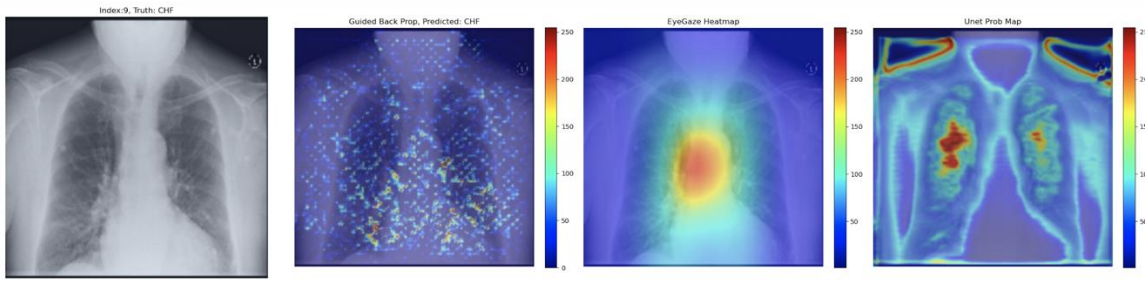

*Figure 41: Model's output heatmaps using guided back-propagation as the heatmap generator for Experiment Set 2 (for "CHF" class correctly classified)*

Appendix C.2 Outputs with deconvNet

Figures 42, 43, and 44 were the outputs of the Experiment Set 2 using deconvNet as the heatmap generator.

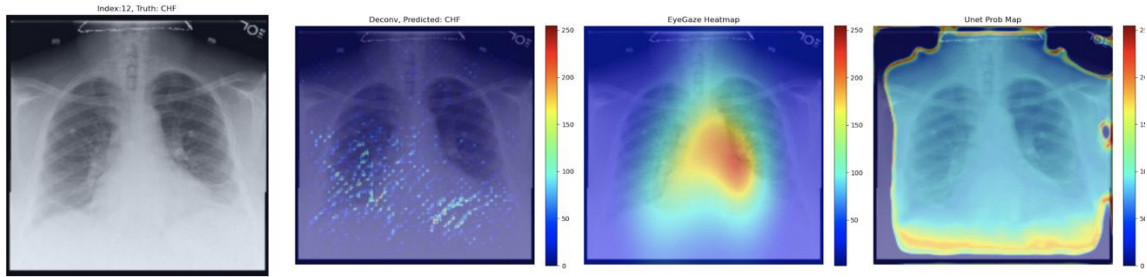

*Figure 42: Model's output heatmaps using deconvNet as the heatmap generator for Experiment Set 2 (for "CHF" class correctly classified)*

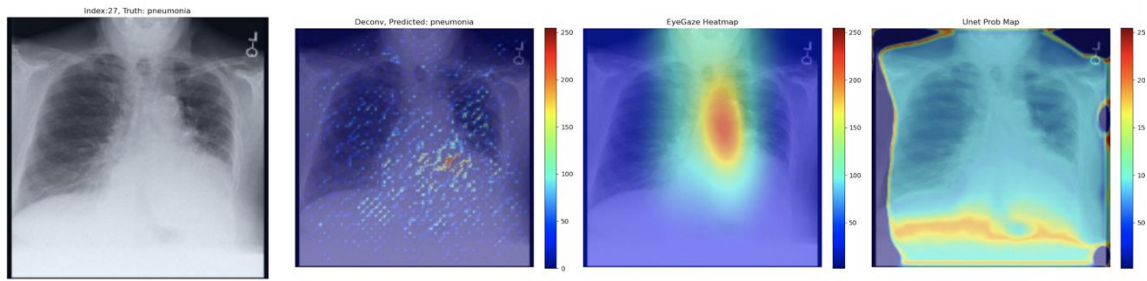

*Figure 43: Model's output heatmaps using deconvNet as the heatmap generator for Experiment Set 2 (for "pneumonia" class correctly classified)*

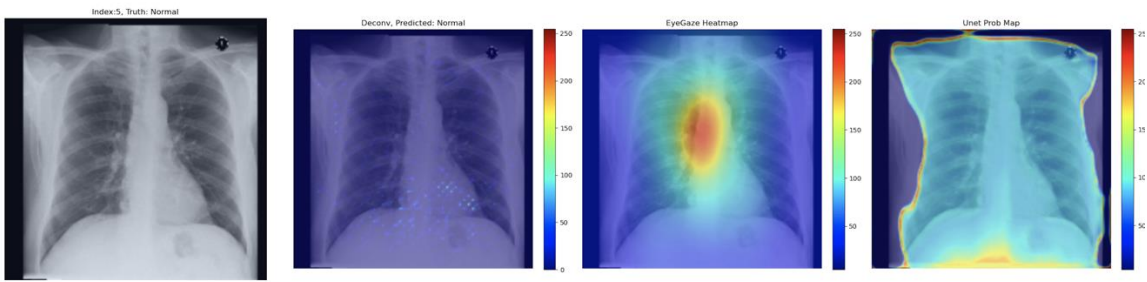

*Figure 44: Model's output heatmaps using deconvNet as the heatmap generator for Experiment Set 2 (for "normal" class correctly classified)*

Appendix D. Outputs of Experiment Set 3

Appendix D.1 Outputs with guided back-propagation

Figures 45, 46, and 47 were the outputs of the Experiment Set 3 using guided back-propagation as the heatmap generator.

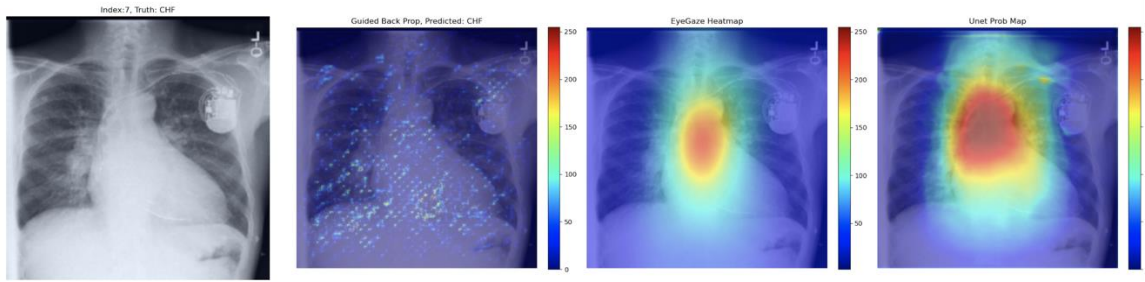

*Figure 45: Model's output heatmaps using guided back-propagation as the heatmap generator for Experiment Set 3 (for "CHF" class correctly classified)*

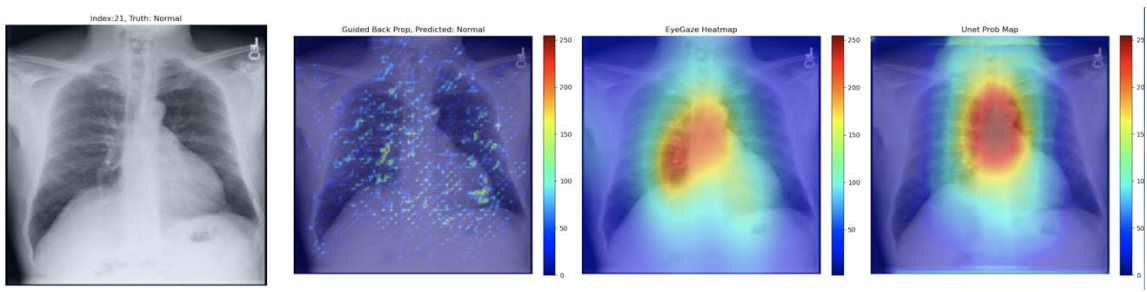

*Figure 46: Model's output heatmaps using guided back-propagation as the heatmap generator for Experiment Set 3 (for "normal" class correctly classified)*

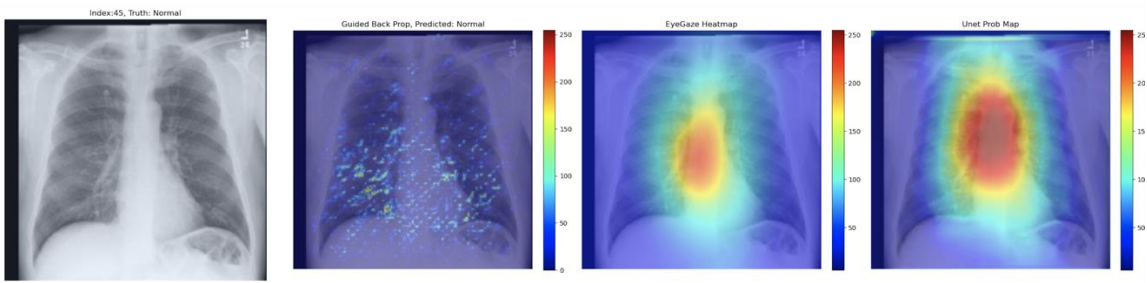

*Figure 47: Model's output heatmaps using guided back-propagation as the heatmap generator for Experiment Set 3 (for "normal" class correctly classified)*

Appendix D.2 Outputs with deconvNet

Figures 48, 49, and 50 were the outputs of the Experiment Set 3 using deconvNet as the heatmap generator.

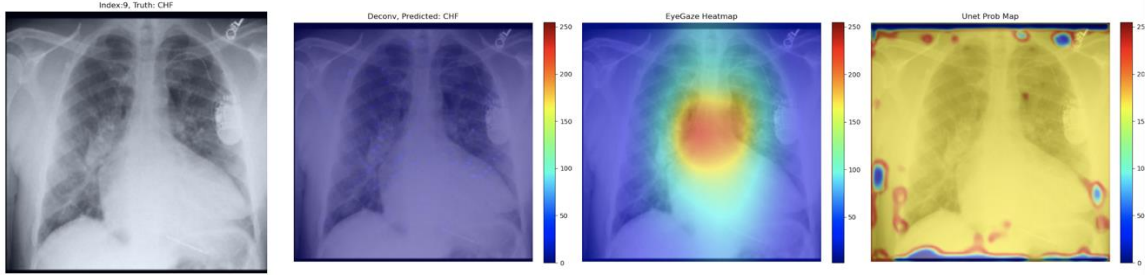

*Figure 48: Model's output heatmaps using deconvNet as the heatmap generator for Experiment Set 3 (for "CHF" class correctly classified)*

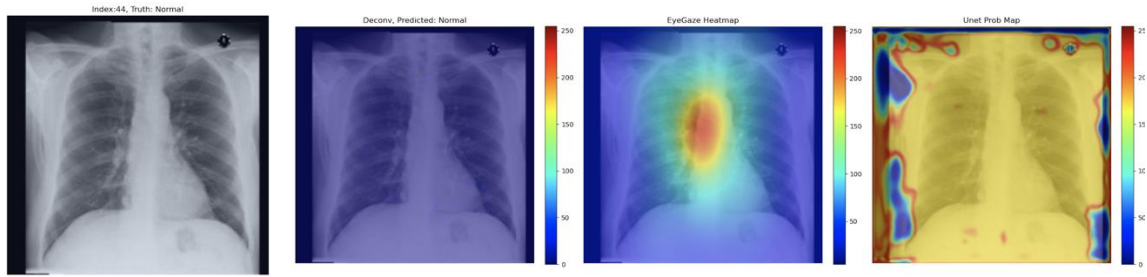

*Figure 49: Model's output heatmaps using deconvNet as the heatmap generator for Experiment Set 3 (for "normal" class correctly classified)*

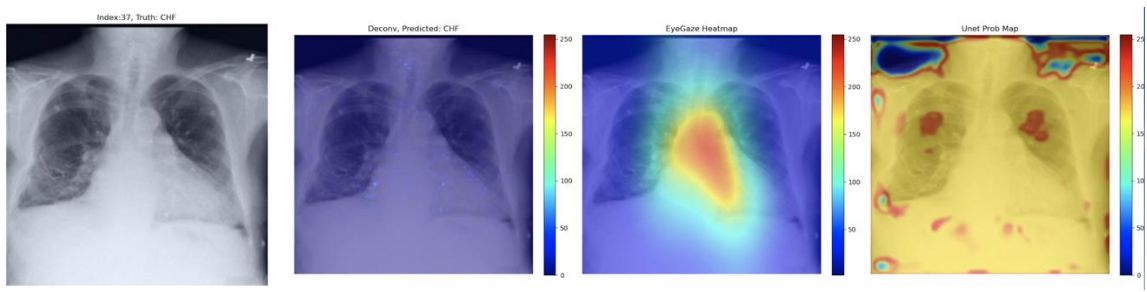

*Figure 50: Model's output heatmaps using deconvNet as the heatmap generator for Experiment Set 3 (for "CHF" class correctly classified)*